\documentclass{jfm}

\usepackage{graphicx}
\usepackage{newtxtext}
\usepackage{newtxmath}
\usepackage{natbib}
\usepackage{hyperref}
\usepackage{booktabs}
\usepackage{multirow}
\usepackage{comment}

\hypersetup{
    colorlinks = true,
    urlcolor   = blue,
    citecolor  = black,
}

\newcommand{\RomanNumeralCaps}[1]
\linenumbers

\title{Interactions Between Internal Solitary Waves and Floating Canopies}

\author{Jen-Ping Chu\aff{1}
  \corresp{\email{jenpingc@usc.edu}},
  Mitul Luhar\aff{1}
 \and Patrick Lynett\aff{2}}

\affiliation{\aff{1}Department of Aerospace and Mechanical Engineering, University of Southern California,
Los Angeles, CA 90010, USA
\aff{2}Department of Civil Engineering, University of Southern California, Los Angeles, CA 90010, USA}

\makeatletter
\gdef\@underjournal{}
\makeatother

\makeatletter
\def\ps@titlepage{\leftskip\z@\let\@mkboth\@gobbletwo\vfuzz=5\p@
  \def\@oddhead{\hbox to \textwidth{\@j@urnal \hfil\llap{\thepage}}}
}
\makeatother

\makeatletter
\def\pagelimitfooter{}
\makeatother

\begin{document}
\maketitle

\begin{abstract}
Interactions between internal solitary waves and floating canopies of varying length and porosity are examined via laboratory experiments and complementary simulations for a miscible, two-layer system. In both approaches, internal solitary waves of varying amplitudes are generated by a jet-array mechanism that is driven by the nonlinear eKdV solution. Pycnocline displacements, phase speeds, and velocity fields are obtained using synchronized planar laser-induced fluorescence and particle imaging velocimetry systems in the experiment. In the simulations, the canopy is represented as a porous zone with prescribed porosity and hydraulic conductivity determined by the Kozeny–Carman model, which is validated by comparing simulated and measured horizontal velocity profiles. The higher-porosity (transitional) canopy produces a nearly monotonic, albeit minor, amplitude reduction and negligible wave energy dissipation after the interaction. However, the shear layer developed at the bottom edge of the lower-porosity (dense) canopy grows to a comparable strength as the shear sustained by the internal solitary wave profile at the pycnocline. The vortex pair generated by this shear accelerates the upper-layer fluid beneath the canopy, leading to complex nonlinear amplitude modulation and significant wave transformation. With an extended canopy length, the internal solitary waves settle to a quasi-steady state with a significant phase speed reduction. Upon the wave exiting the canopy, flow separation at the downstream edge of the canopy again pairs with the shear at the pycnocline, inducing an intensified jet.   This complex interaction leads to energy transfer between kinetic  and potential energy under the dense canopy. 

\end{abstract}



\section{Introduction}
\label{sec:intro}
Variations in seawater temperature and salinity levels establish ocean stratification. Baroclinic forcings within such stratified fluid systems - for example, tidal flow over bottom topography or opposing background currents - evolve into internal waves (IWs) propagating along pycnoclines. One class of IWs - internal solitary waves (ISWs) - can maintain a localized, single-hump structure at the pycnocline over an extended distance through a balance between nonlinear steepening and linear dispersion \citep{Helfrich2006, Cavalier2021}. The occurrence of ISWs is well documented through both in-situ thermistor array and synthetic aperture radar (SAR) imagery. ISWs have been observed on continental shelves, in straits, and marginal seas \citep{Apel2022, Duda2004}. As they propagate, ISWs transport energy as well as biological and chemical constituents. The interaction between ISWs, bathymetry, and ambient shear triggers vertical mixing and plays a key role in biogeochemical processes. For example, mixing often introduces bottom nutrients into the upper water column, thereby fertilizing the local region and modifying primary production and related biological processes.

Near-surface floating obstructions, such as naturally occurring macroalgae mats (e.g., the Sargasso Sea), macroalgae farms, engineered offshore platforms, and sea ice, have the potential to modify the propagation of ISWs. The existence of these obstructions may therefore alter the sustained propagation of ISWs as well as interfacial mixing processes. 
 
Previous investigations of the hydrodynamic interactions between ISWs and floating ice have reported intensified flow and interfacial overturning downstream of the ice obstruction, accompanied by enhanced turbulent dissipation \citep{Zhang2022, Terletska2024, Hartharn2024}. Upon impact, the incident ISW disintegrates and radiates its energy into smaller secondary waves. Different regimes of energy transformation emerge depending on the level of obstruction. A solid obstruction thinner than the upper layer depth of a two-layer system favors energy transmission with dissipation, whereas an obstruction thicker than the upper layer depth promotes reflection over transmission. 
While studies on the interactions between ISWs and floating solid obstructions have increased in the past five years, relatively less attention has been paid to the interaction between ISWs and floating porous objects, as will be presented herein. The porous canopy structure allows partial flow penetration. Canopy-scale vortices within the porous structure and the shear at the canopy interface are expected to augment dissipation.  In addition to providing insight into the interaction between ISWs and floating structures, this study can inform offshore macroalgal farming, proposed as a sustainable strategy for greenhouse gas absorption, biofuel production, and food supply \citep{Frieder2022, Arzeno2023}. Understanding the fundamental dynamics of ISW-canopy interaction provides a foundation for predicting the impacts and benefits to coastal processes.  

Prior studies have provided substantial insight into the interaction between submerged canopies and unidirectional or wave-induced flows.  For a unidirectional flow within an unstratified fluid, the velocity profile remains quasi-logarithmic for sparse submerged canopies. For a dense submerged canopy, the drag discontinuity at the canopy tip generates a shear layer, which can induce canopy-scale vortices that control the exchange within the canopy and overflow \citep{Nepf2012}. For progressive wave forcing within an unstratified fluid, a mean current in the direction of wave propagation is generated within the submerged canopy by a nonzero wave stress \citep{Luhar2010}. The mean current can promote the net horizontal transport of suspended sediment and organic matter within the canopy structure. In some ways, floating canopies may be regarded as the inverted analogue of submerged canopies with similar behaviors. However, the free-stream flow bounded by the bottom edge of the floating canopy and the solid bed confines the flow development. The no-slip bottom wall boundary imposes a stronger physical restriction on the growth of the shear layer, which is responsible for generating the coherent vortices at the bottom edge of the floating canopy \citep{Plew2011}. With the additional background stratification, structures resembling Kelvin–Helmholtz (KH) vortices emerge as the shear layer at the bottom edge of the canopy interacts with the underlying stratification, radiating small-scale internal waves \citep{Chamecki2025}. Sufficiently strong stratification is shown to suppress vertical mixing, favoring a horizontal motion instead, and thereby enhancing the dissipation within the canopy structure \citep{Plew2006}.

In this paper, the effect of a floating canopy on internal solitary waves is examined experimentally and numerically in a system of miscible two-layer stratification. This two-layer system creates stronger confinement between the bottom of the canopy and the pycnocline, which can lead to enhanced interactions between the shear layers generated by the ISWs at the bottom of the canopy and at the pycnocline. Through experimental and numerical efforts, we characterize the influence of canopy porosity and length on the propagation of internal solitary waves of varying amplitudes. Section \ref{sec:exp} describes the experimental configuration, measurement techniques, and the explored parameter maps. Section \ref{sec:num} describes the complementary numerical modeling effort, including the permeability specification for the canopy. Section \ref{sec:res} presents amplitude evolution during the interaction, the corresponding mechanisms at different interacting phases, and the associated energy transformation. Finally, concluding remarks are offered in Section \ref{sec:conclusion}. An additional simulation-based discussion of extended canopy length in order to isolate transient effects is provided in Appendix \ref{appA}.

\section{Experimental Method}
\label{sec:exp}

\subsection{Experimental Setup}
\label{sec:exp_setup}
The ISWs were generated by a jet-array wave maker (JAW) \citep{Chu2025} in a $2.2$ m long, $0.2$ m wide, and $0.5$ m deep acrylic flume. Within the flume, a miscible two-layer system was set up as background stratification, separated by a thin pycnocline. The lower layer was filled with the salt solution of density, $\rho_2 = 1020 \pm 2$ kg/m$^3$. This was confirmed via multiple measurements using a salinity probe. Freshwater with density of $\rho_1 = 998$ kg/m$^3$ was then introduced into the flume until the total depth $H = h_1 + h_2$ reached approximately $11$ cm. The experiments were carried out with the fixed nominal depth ratio of $h_2/h_1 = 9/2$ with nominal wave amplitudes varying from $a = (-1, \, -1.5, \, -2)$ cm. 

\begin{figure}
    \centerline{\includegraphics[width = 1.0\textwidth]{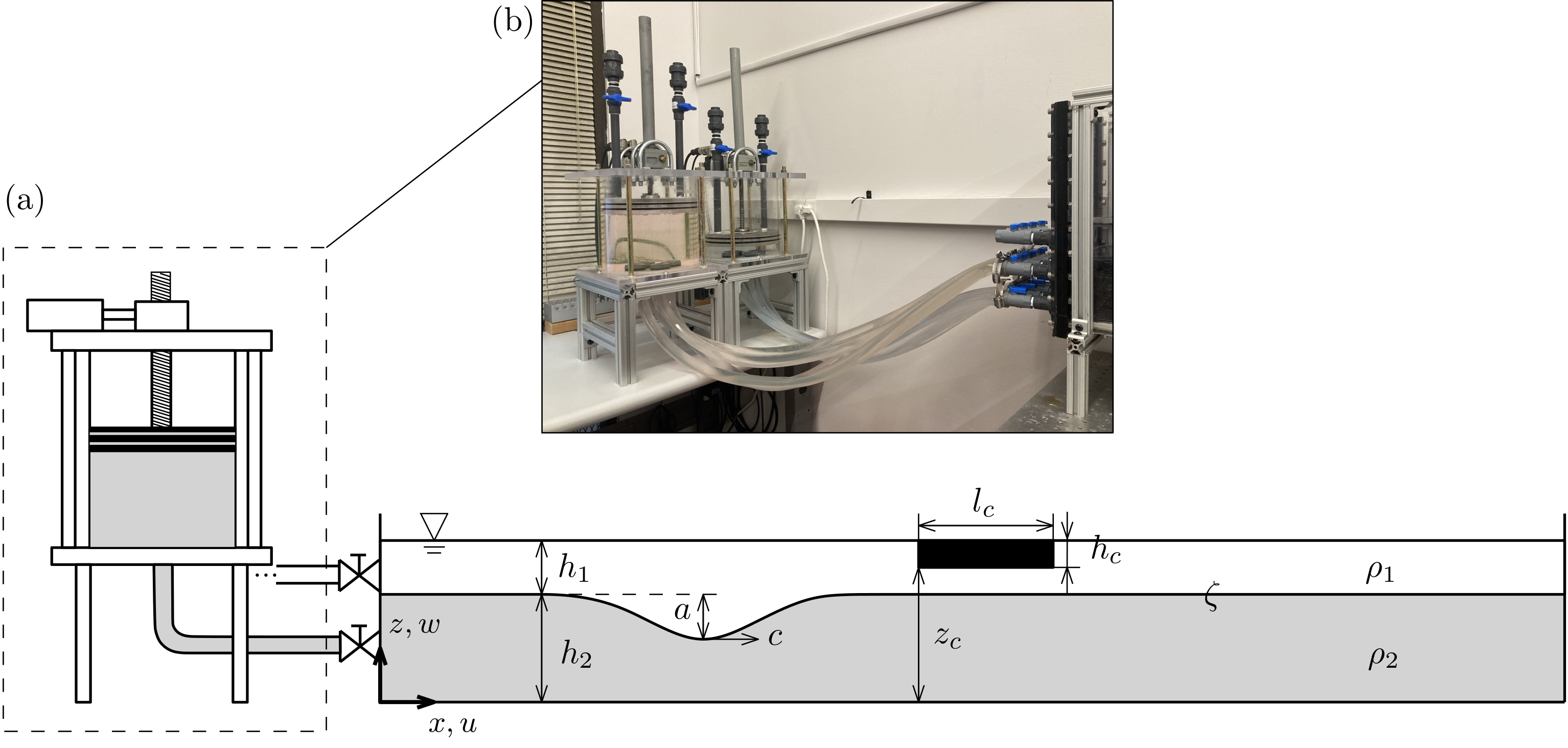}}
    \caption{(a) Schematic for the experimental set-up of interactions between ISWs and floating canopies, (b) a photo image of the jet-array wave maker.}
    \label{fig:setup}
\end{figure}

The JAW system is composed of two independent, piston-driven cylinders, as provided in Figure \ref{fig:setup} (b). The cylinders containing freshwater and salt solution are connected to the upper and lower layers of the flume, respectively. The pistons were driven by servo motors (Kollmorgen AKM44J), which were controlled using MATLAB scripts interfaced with a National Instruments data acquisition system (NI 6009) to generate prescribed volumetric fluxes (inflows and outflows) within each layer.  To ensure near-uniform flow within each layer, fluid was driven from each cylinder through four uniformly spaced pipes, which subsequently fed through honeycomb flow straighteners. For each trial, the signal to drive the JAW system was prescribed based on the extended Korteweg de Vries (eKdV) solution using the nominal layer depths and amplitude. The influence of a slight variation between nominal and actual layer depths on wave generation using the JAW system is addressed in \citet{Chu2025}.  For the conditions considered in this paper, the actual layer depths differed from the nominal values by $<0.2$ cm. This is expected to translate into wave amplitude differences of 10 \% or less relative to the nominal prescribed values.

\begin{figure}
    \centerline{\includegraphics[width = 0.9\textwidth]{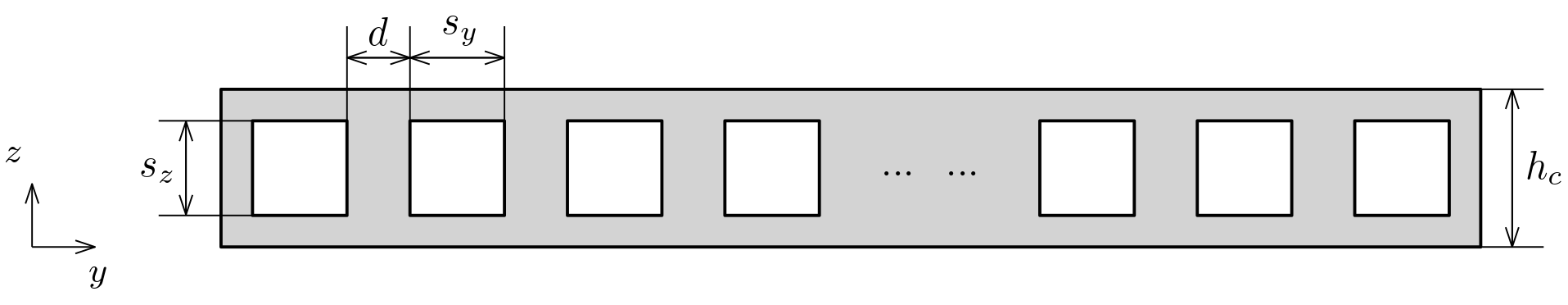}}
    \caption{Cross-sectional view and nomenclature of the canopy structure. }
    \label{fig:canopy}
\end{figure}

Floating canopy structures were installed approximately $100$ cm downstream of the inlet, positioned just beneath the free surface at a vertical location of $z_c = 10$ cm as shown in Figure \ref{fig:setup} (a). The canopies were 3D-printed to span the full channel width, with a constant height of $h_c = 1$ cm. Each canopy consisted of uniform element width, $d$, equal streamwise and lateral spacing of $s_x = s_y$, and a vertical spacing of $s_z$ as illustrated in Figure \ref{fig:canopy}. Two canopy lengths ($l_c = 5h_1 = 10$ cm and $l_c = 10h_1 = 20$ cm) and two porosity levels( $n = 0.648, 0.964$) were tested. The printed dimensions were $(d, s_x=s_y, s_z) = (0.4, 0.6, 0.6)$ cm for the lower porosity canopies $(n=0.648)$, and $(d, s_x=s_y, s_z) = (0.2, 2.3, 0.8)$ cm for the higher porosity canopies $(n=0.964)$. 

These two porosities were chosen to span the so-called dense and transitional regimes for canopy flows \citep{Nepf2012}. Previous work shows that flow penetration into canopies from adjacent non-vegetated regions depends on the dimensionless frontal area ratio $\lambda_f$, which represents the frontal area of the canopy per unit of planar area in the streamwise direction.  For $\lambda_f \ll 0.1$, the canopy can be considered sparse, and there is significant flow penetration into the canopy.  For $\lambda_f \gg 0.1$, the canopy can be considered dense.  In this "dense" case, a region of strong shear develops at the interface between the canopy and the adjacent unobstructed flow. This canopy shear layer can generate large-scale turbulent rollers via mechanisms similar to the Kelvin-Helmholtz instability.  Canopies with $\lambda_f \approx 0.1$ exhibit transitional behavior between the sparse and dense regimes. For the floating canopy geometries considered in this paper, the less porous canopy has a frontal area ratio of $\lambda_f \approx 0.64$ and the more porous canopy has a frontal area ratio of $\lambda_f \approx 0.16$.  These canopies may therefore be classified as dense and transitional, respectively.

Table \ref{tab:case_exp} summarizes the parameters for the $12$ experiments conducted. Each case is named after the prescribed wave amplitude (A), canopy length (L), and canopy type — dense (D) or transitional (T) — corresponding to its porosity. In this table, $h_{1,m}$ and $h_{2,m}$ are the measured layer depths obtained using planar laser-induced fluorescence, as described below.  

\begin{figure}
    \centerline{\includegraphics[width = 0.7\textwidth]{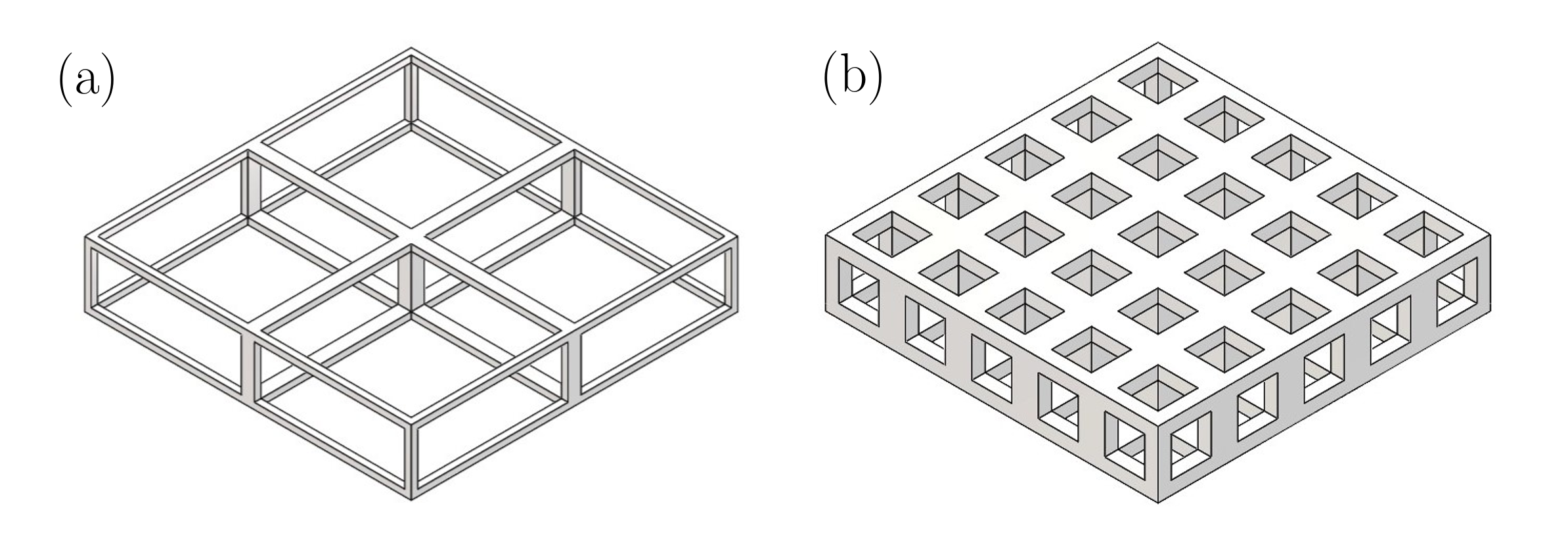}}
    \caption{Isometric views of the canopy structures. (a) Transitional canopy with porosity $n=0.964$. (b) Dense canopy with $n=0.648$.}
    \label{fig:sample}
\end{figure}

\begin{table}
\centering
\begin{tabular}{l c c c c c c}
\toprule
Case & $h_{1,m}$ (cm) & $h_{2,m}$ (cm) & $a$ (cm) & $l_c$ (cm) & $n$ &  Marker\\
\midrule
A1L1T & 2.23 & 8.93 & 1.0 & 10.0 & 0.964 & {\color{black} \footnotesize $\,\,\triangle$} \\
A1.5L1T & 2.18 & 9.18 & 1.5 & 10.0 & 0.964 & {\color{blue} \footnotesize $\,\,\triangle$} \\
A2L1T & 2.15 & 9.03 & 2.0 & 10.0 & 0.964 & {\color{red} \footnotesize $\,\,\triangle$} \\
A1L2T & 2.17 & 9.05 & 1.0 & 20.0 & 0.964 & {\color{black} \normalsize $\,\,\triangledown$} \\
A1.5L2T & 2.12 & 8.93 & 1.5 & 20.0 & 0.964 & {\color{blue} \normalsize $\,\,\triangledown$} \\
A2L2T & 2.16 & 9.06 & 2.0 & 20.0 & 0.964 & {\color{red} \normalsize $\,\,\triangledown$} \\
A1L1D & 2.05 & 9.18 & 1.0 & 10.0 & 0.648 & {\color{black} \footnotesize $\,\,\square$} \\
A1.5L1D & 2.11 & 9.12 & 1.5 & 10.0 & 0.648 & {\color{blue} \footnotesize $\,\,\square$} \\
A2L1D & 2.11 & 9.05 & 2.0 & 10.0 & 0.648 & {\color{red} \footnotesize $\,\,\square$} \\
A1L2D & 2.11 & 9.05 & 1.0 & 20.0 & 0.648 & {\color{black} \normalsize $\,\,\lozenge$} \\
A1.5L2D & 2.17 & 9.17 & 1.5 & 20.0 & 0.648 & {\color{blue} \normalsize $\,\,\lozenge$} \\
A2L2D & 2.18 & 9.12 & 2.0 & 20.0 & 0.648 & {\color{red} \normalsize $\,\,\lozenge$} \\
\bottomrule
\end{tabular}
\caption{Theoretical and measured values used for the experiments.}
\label{tab:case_exp}
\end{table}

\subsection{Measurement Techniques}
\label{sec:meas}
\begin{figure}
    \centerline{\includegraphics[width = 0.5\textwidth]{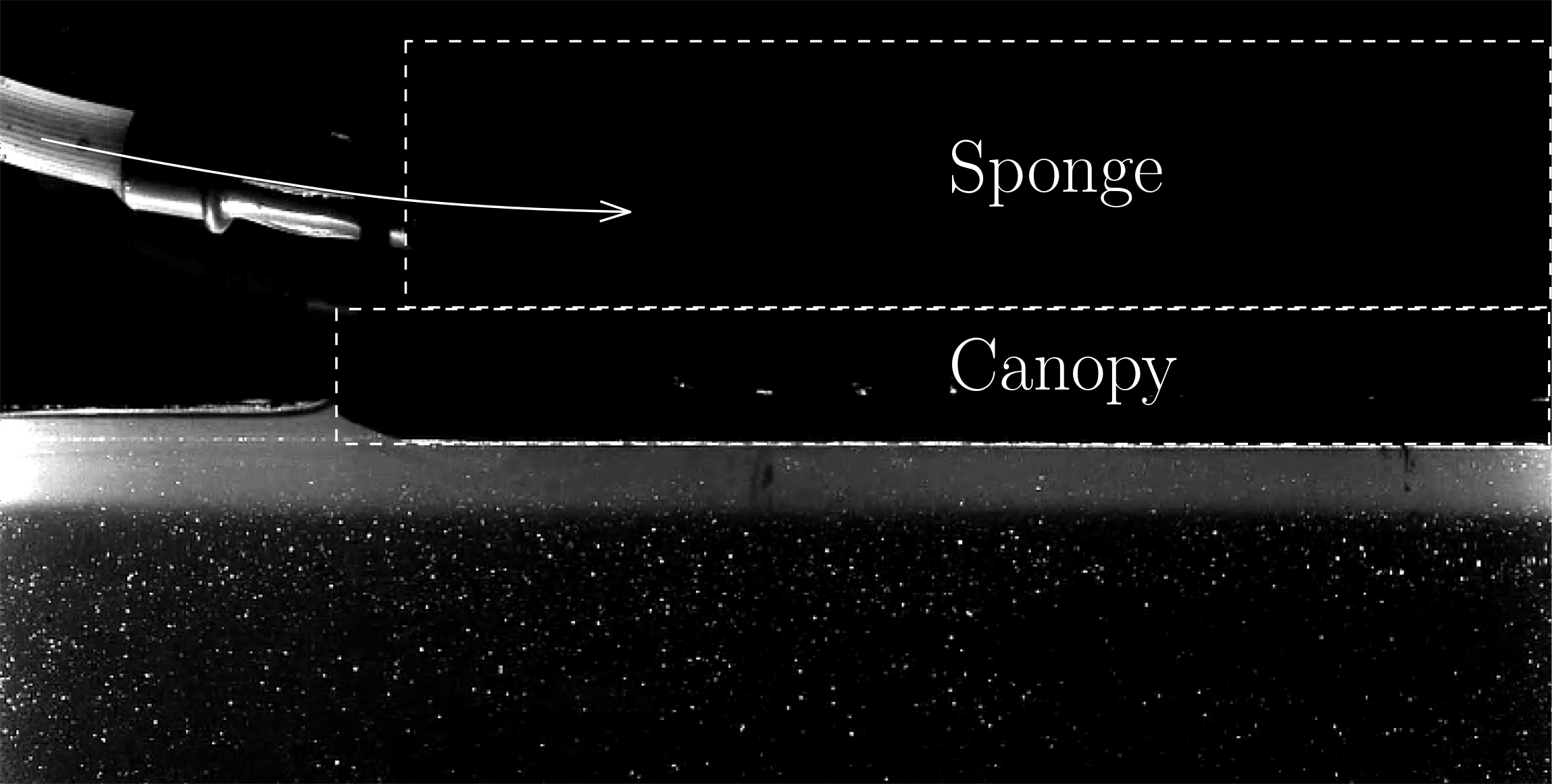}}
    \caption{An illustration of the two-step filling method to ensure sufficient particle density in the upper layer.}
    \label{fig:seeding}
\end{figure}

Interactions between ISWs and canopies were captured by synchronized planar laser-induced fluorescence (PLIF) and two-dimensional two-component particle imaging velocimetry (2D-2C PIV). Due to particle settling effects, PIV measurements in prior work were subject to insufficient particle density in the uppermost region of the freshwater layer, which is the location of our floating canopy structure. To ensure sufficient seeding density around the canopy, a revised two-step filling method was employed here. First, the freshwater layer was built with sponges floating on the free surface until the water level reached the canopy height. Second, the sponges were repositioned on top of the canopy as illustrated in Figure \ref{fig:seeding} with the seeded fresh water slowly dripping into the upper layer. Due to the sponge repositioning, the pycnocline thickness estimated from PLIF increased to around $1.5$ cm in the experiments considered here (as opposed to remaining below $1$ cm in the previous work).

\begin{figure}
    \centerline{\includegraphics[width = 0.8\textwidth]{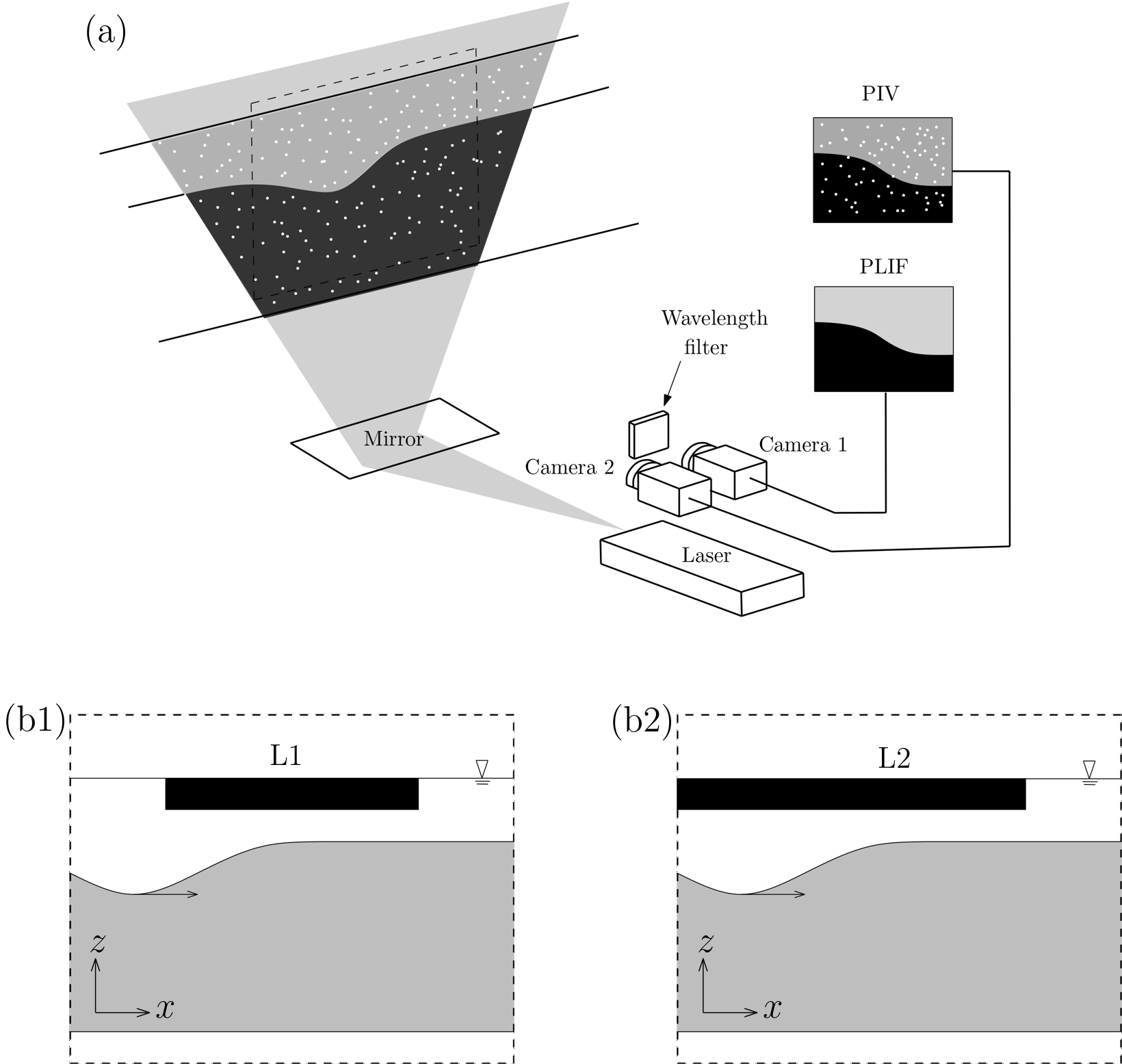}}
    \caption{(a) Schematic of the synchronized PLIF and PIV measurement system. (b1) Observation window coverage for short canopy, (b2) for long canopy conditions.}
    \label{fig:obs}
\end{figure}

The synchronized PLIF and PIV system consisted of two 8-bit CMOS monochrome cameras (Mako-U-130B) of $1024 \times 1280$ pixel resolution operating at $50$ Hz for $40$ s. The measurement duration was chosen to ensure that the entirety of the ISW was captured. A 5W continuous laser with an emission wavelength 532 nm and built-in sheet generation optics was used to illuminate the flow field from below (Figure \ref{fig:obs} (a)). For the PLIF measurements, Rhodamine 6G dye with a peak excitation wavelength of 525nm and a peak emission wavelength of 548nm was premixed with the freshwater.  An optical filter with a passband between 536 nm to 564 nm was therefore used to isolate the fluorescent freshwater layer. Assuming single-pixel resolution for the PLIF measurements, the dye field is resolved to better than 0.2 mm.

The cameras shared an overlapping field-of-view measuring $12$ cm $\times$ $15$ cm. For the short canopy case $(l_c/h_1 = 5)$, the field-of-view extended from upstream to downstream of the canopy, enabling full observation of the interaction process, as illustrated in Figure \ref{fig:obs} (b1). However, for the long canopy case $(l_c/h_1 = 10)$, the field-of-view could not cover the entire canopy length. As such, the field of view was positioned to focus on the downstream half of the canopy.  Specifically, the field-of-view encompassed slightly more than the downstream half of the canopy with an additional $2$ cm beyond, as illustrated in Figure \ref{fig:obs} (b2). The MATLAB PIVLab package was used to post-process the images using a multi-pass algorithm. The initial pass involved a $64$ pixel $\times$ $64$ pixel interrogation window. Two subsequent passes with $32$ pixel $\times$ $32$ pixel interrogation windows with $50$ \% overlap yielded $52$ $\times$ $32$ velocity vectors per snapshot.

\section{Numerical Model}
\label{sec:num}

\subsection{Model Setup}
\label{sec:model}
Numerical simulations of ISWs interactions with floating canopies were conducted using ANSYS Fluent. A 2D computational domain was designed to match the experimental setup illustrated in the previous section, consisting of a total depth $H = 0.11$ m and a flume length $L=2.2$ m. The domain was discretized using a uniform hexahedral mesh with resolution of $\Delta x = 2\times10^{-3}$ m and $\Delta z = 1\times10^{-3}$ m. Grid resolution in the vertical direction $(z)$ was selected to be no greater than 10\% of the smallest wave amplitude. Grid resolution in the streamwise $(x)$ direction was controlled by the grid aspect ratio ($\Delta x / \Delta z$) to avoid amplified numerical dissipation. To replicate the experimental condition, the initial stratification was imposed using a hyperbolic tangent density profile with a fixed pycnocline thickness of 1.5 cm. The upper freshwater layer was assigned a density of $\rho_1 = 998$ kg/m$^3$ and a dynamic viscosity of $\mu_1 = 1.00 \times 10^{-3}$ Pa$\cdot$s, while the lower salt solution layer was defined by $\rho_2 = 1020$ kg/m$^3$ and a dynamic viscosity of  $\mu_2 = 1.09 \times 10^{-3}$ Pa$\cdot$s.

\begin{figure}
    \centerline{\includegraphics[width = 0.8\textwidth]{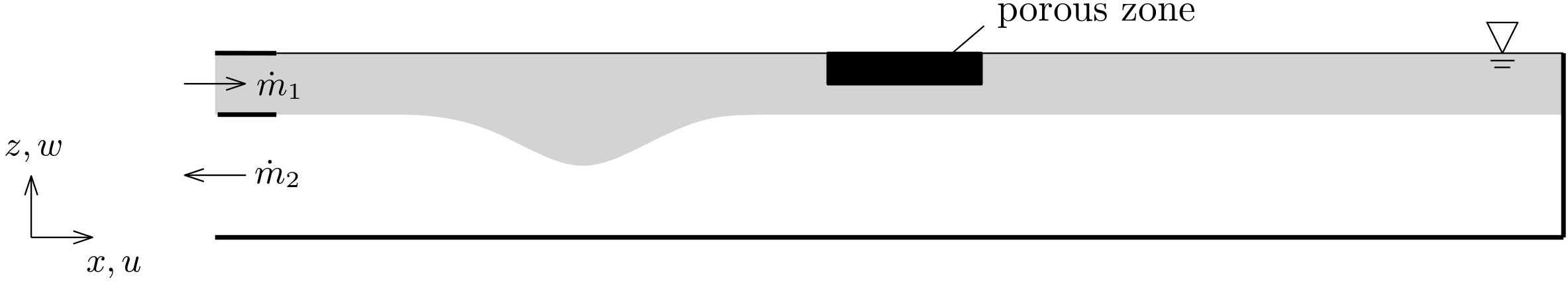}}
    \caption{Schematic showing the numerical model.}
    \label{fig:num_setup}
\end{figure}

To reproduce the wave generation mechanism of the JAW system, mass flow inlets were applied at the upstream boundary of both layers, as illustrated in Figure~\ref{fig:num_setup}. The upper channel allowed only freshwater ($\dot{m_1}$), while the lower channel allowed only ocean water ($\dot{m_2}$). The mass flow rates were calculated by integrating the horizontal velocity profiles from the eKdV solution across the respective layer thicknesses, multiplied by the corresponding fluid density. No-slip boundary conditions were imposed along the channel walls and flume bed. The top surface was modeled as a no-shear boundary to approximate the free surface.

A mixture model was employed to simulate the multiphase configuration of the two-layer stratified system. Turbulence was modeled using the realizable k-$\epsilon$ Reynolds-averaged Navier–Stokes (RANS) model. To ensure accurate reproduction of wave amplitude evolution, the model coefficients within the k-$\epsilon$ formulation were calibrated to avoid overprediction of numerical dissipation relative to experimental measurements. Specifically, the coefficients were set as $c_{2,\epsilon} = 1.6, \sigma_k  =0.8$, and $\sigma_\epsilon =0.9$. Numerical solution procedures employed the coupled pressure-velocity scheme combined with the PRESTO pressure discretization. Second-order implicit upwind schemes were applied to all transport equations, including momentum, turbulent kinetic energy, dissipation rate, and transient formulation. A fixed time step of $ \Delta t = 5 \times 10^{-3}$ s was employed throughout all simulations. This value was selected to satisfy the Courant–Friedrichs–Lewy (CFL) stability criteria. The CFL number was estimated based on the dominant streamwise propagation of an ISW of $a = 2$ cm by using the nominal phase speed, $c$, as the upper limit for the horizontal particle speed. This estimation resulted in a maximum CFL $= (c \Delta t)/ \Delta x \approx 0.18$.

\begin{figure}
    \centerline{\includegraphics[width = 0.9\textwidth]{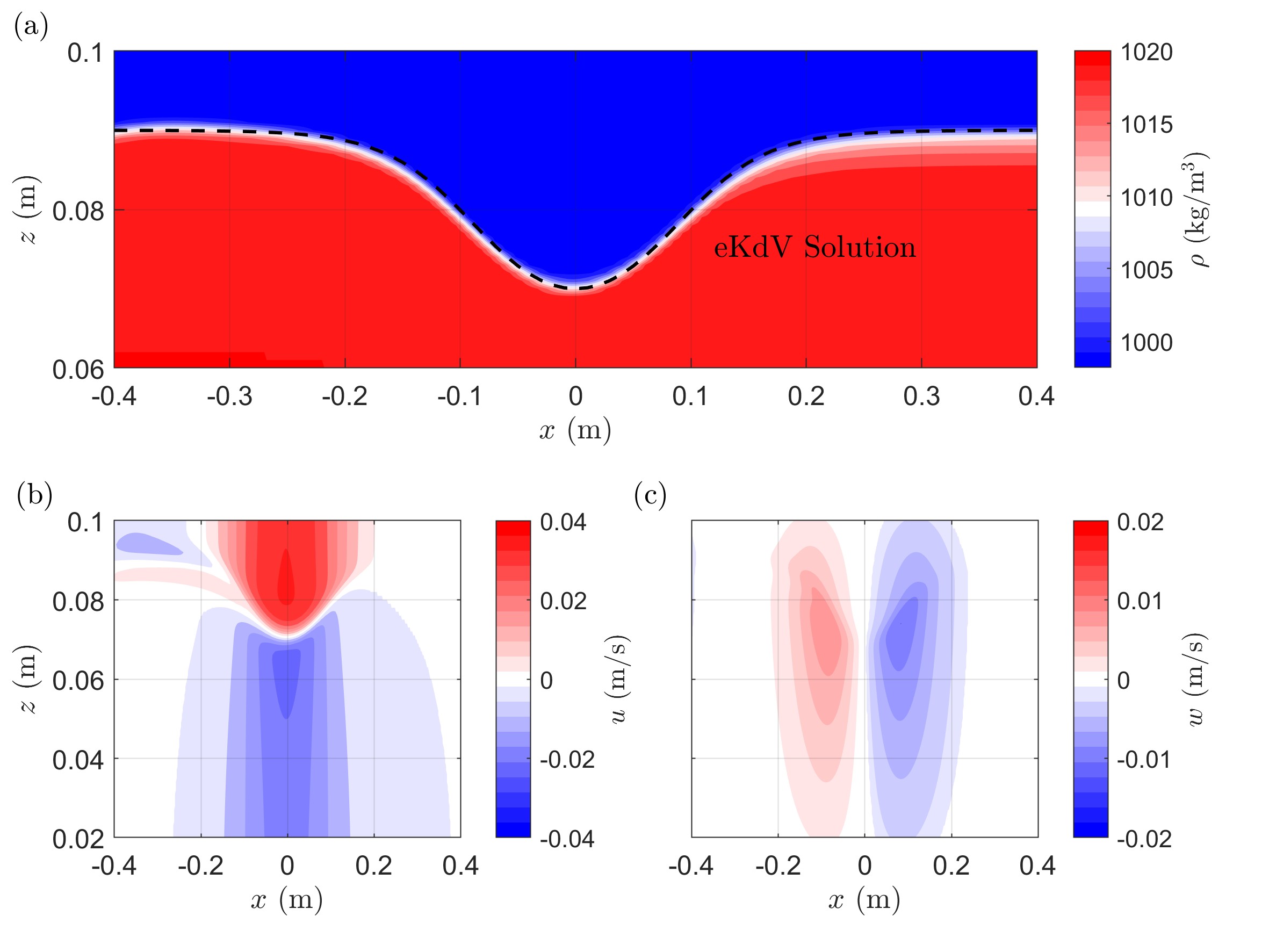}}
    \caption{Contours showing (a) density, (b) horizontal velocity and (c) vertical velocity for a simulated ISW. The black dashed line in (a) demonstrates the eKdV solution of the same ISW.}
    \label{fig:num_verf}
\end{figure}

\begin{figure}
    \centerline{\includegraphics[width = 1.0\textwidth]{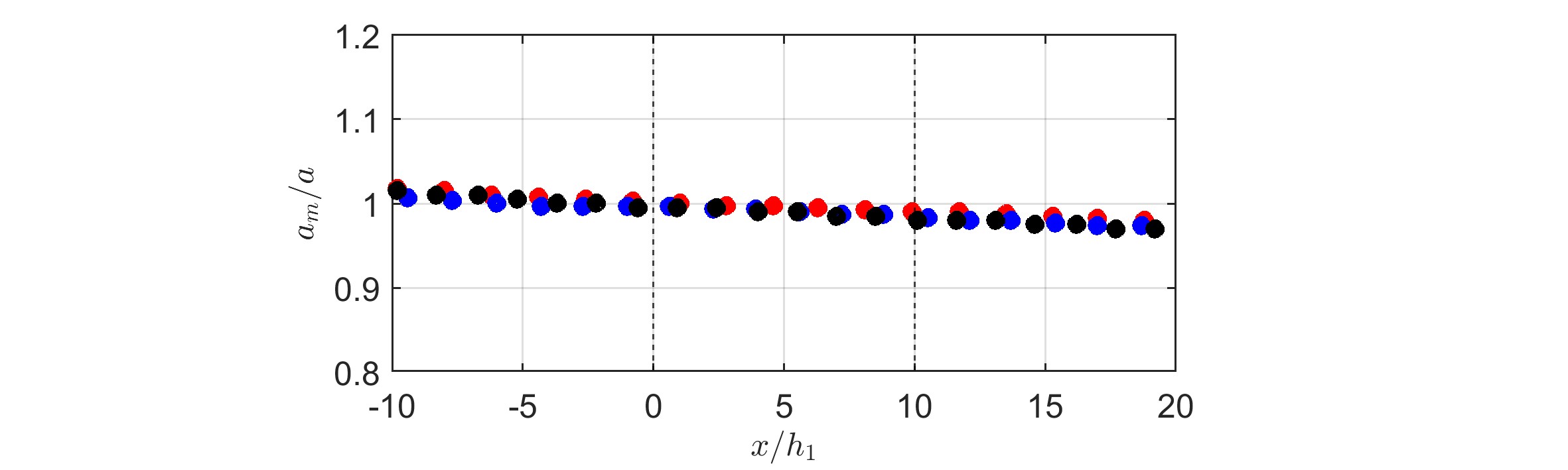}}
    \caption{Spatial variation of ISW amplitude, where $a_m$ denotes the simulated value and $a$ represents the theoretical value. The vertical dashed lines show the extent of the long canopies. Black, blue, and red markers correspond to ISW amplitudes $a = \{1, 1.5 ,2\}$ cm.}
    \label{fig:cfd_L0_x_a}
\end{figure}

Figure~\ref{fig:num_verf} presents a snapshot of density, horizontal velocity, and vertical velocity contours of a simulated ISW corresponding to $(h_1, h_2, a) = (2,9,2)$ cm, evaluated at $1$ m downstream from the inlet. For consistency in the subsequent analysis, the streamwise coordinate origin $(x=0)$ is defined at $1$m downstream from the inlet, where the leading edge of the floating canopies would otherwise be. The wave profile of the generated ISW is closely aligned with the eKdV solution for the same layer depths and wave amplitude. The spatial evolution of ISWs amplitudes is illustrated in Figure~\ref{fig:cfd_L0_x_a}. In the absence of a canopy, the ISW exhibited an amplitude reduction of approximately $0.5 - 1\%$ within the region that would otherwise be canopy-covered, and a total amplitude loss of $2 - 3\%$ over a distance of approximately $30h_1$.  

\subsection{Canopy Zone Characterization}
\label{sec:perm}
The floating canopy structure was modeled as a homogeneous porous zone within the computational domain, as illustrated in Figure~\ref{fig:num_setup}. The porous region was defined by a specified porosity, $n$, and a viscous resistance parameter, $1/K$, representing the inverse of permeability ($K$). While the porosity was directly calculated as listed in Table~\ref{tab:case_exp}, accurate determination of viscous resistance was more challenging in experiments due to the small measurable pressure gradient across our canopy dimensions.

Previous experimental studies by \citet{Shilpa2024} characterized discrete viscous resistance values for both isotropic and anisotropic porous substrates with cubic lattice geometries similar to those considered in this paper.  However, no general predictive relationship was established for estimating the viscous resistance of these porous structures. In the present study, the Kozeny–Carman equation (Eq.~\ref{eq:KC}) was adopted as a base model for estimating permeability \citep{Carman1939}:
\begin{equation}
    K_{KC} = \frac{\beta}{\sigma^2} \frac{n^3}{1-n^2}.
    \label{eq:KC}
\end{equation}
In this model, $\sigma$ represents the specific surface area for the structure, and $\beta$ is the so-called Kozeny constant, which depends on the specific porous geometry. The Kozeny-Carman (KC) model was initially derived under the assumption of non-interconnected, parallel tube bundles as a simplified representation of complicated porous formations. More recent findings by \citet{Schulz2019} indicate that a value of $\beta=0.2$ yields the best agreement with experimental data for irregular porous structures. The specific surface is area defined as the surface area, $S_c$, divided by the solid volume, $V_c$, of a unit canopy cell. For the geometries considered in this paper, the unit canopy cell consists of a central cube with side length $d$, and three square columns oriented along the principal axes as shown in Figure~\ref{fig:sigma_unit}. As such, the specific surface area can be estimated using Eq.~\ref{eq:spec_surf} below:
\begin{equation}
    \sigma = \frac{S_c}{V_c} = \frac{6d^2 + 4 s_x d + 4 s_y d + 4 s_z d}{d^3 + d^2s_x + d^2 s_y + d^2 s_z}.
    \label{eq:spec_surf}
\end{equation}

\begin{figure}
    \centerline{\includegraphics[width = 0.5\textwidth]{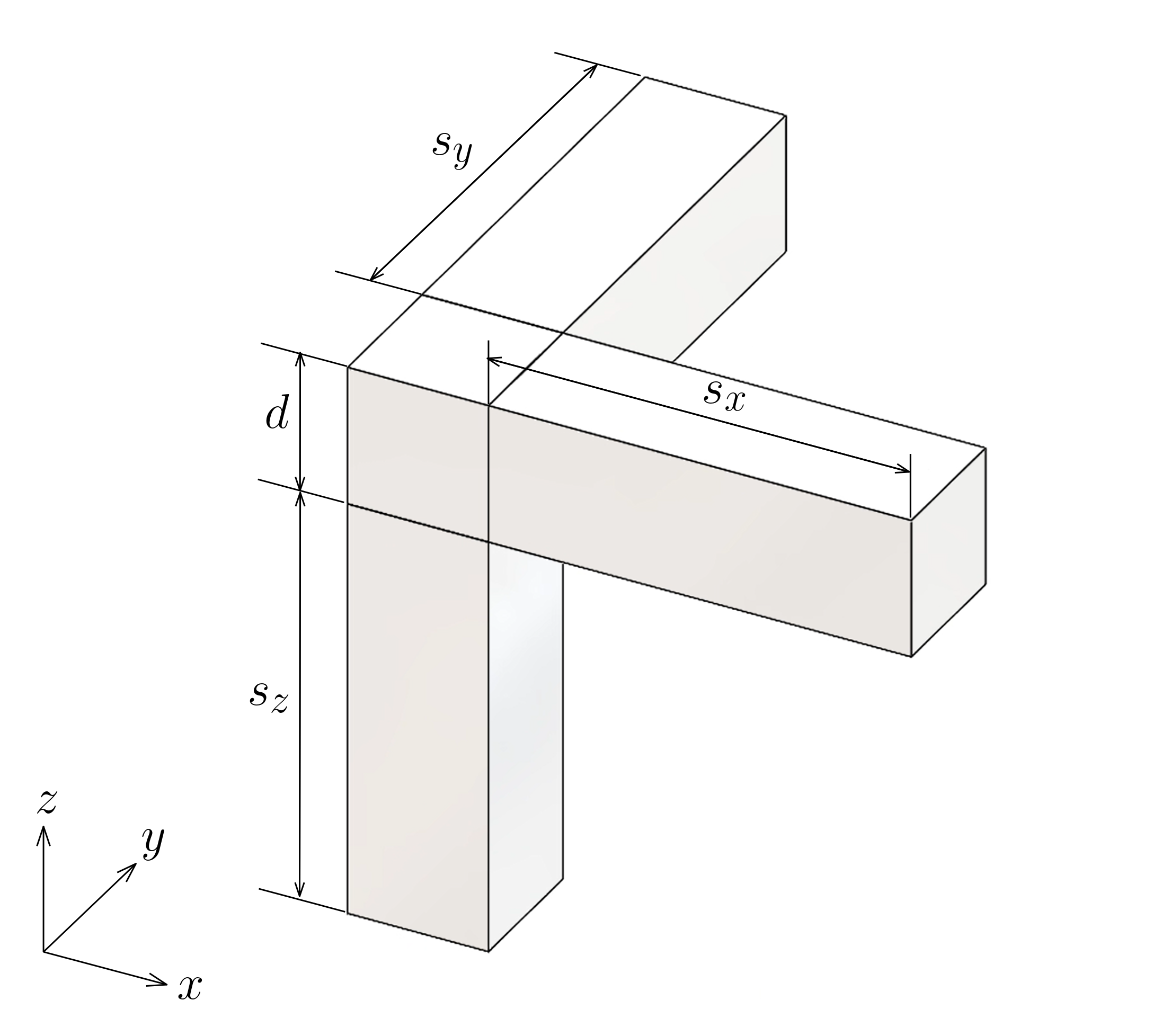}}
    \caption{Unit cell for the calculation of specific surface.}
    \label{fig:sigma_unit}
\end{figure}

To evaluate the applicability of the KC model for the cubic lattice geometries considered in this paper, the predicted resistance values were compared against the measurements reported by \citet{Shilpa2024}, as shown in Table~\ref{tab:perm_com}.  The experiments of \citet{Shilpa2024} considered cubic cell geometries with a fixed stem thickness of $d= 4\times10^{-3}$ m and varying spacings in the principal directions. Each specimen was labeled with a three-letter code representing the spacings along the principal directions.  Here, we consider the isotropic cases HHH, MMM, and LLL, denoting high, medium, and low spacings (i.e., going from more porous to less porous). As shown in Table~\ref{tab:perm_com}, the KC model provides a reasonable estimate of the measured viscous resistances, with predictions agreeing with the measurements within $\pm 15\%$ for the HHH and MMM cases.  For the densest $LLL$ case considered by \citet{Shilpa2024}, the predicted viscous resistance is a factor of 2-3 higher.  Following this limited validation exercise, additional numerical sensitivity analyses were carried out to identify the most appropriate resistance values for the floating canopies considered in the experiments.  

\begin{table}
\centering
\begin{tabular}{l c c c c c c c}
\toprule
Specimen & $d$ (m) & $s_x$ (m) & $s_z$ (m) & $n$ & $\sigma$ ($\mathrm{m^{-1}}$) & $\frac{1}{K_{KC}}$ ($\mathrm{m^{-2}}$) & $\frac{1}{K_{m}}$ ($\mathrm{m^{-2}}$) \\
\midrule     
(\textit{Present Study}) & & & & & & & \\
Transitional Canopy & $2.0 \times 10^{-3}$ & $2.3 \times 10^{-2}$ & $8.0 \times 10^{-3}$ & 0.964 & 0.20 & $2.97 \times 10^{4}$ & - \\
Dense Canopy & $4.0 \times 10^{-3}$ & $6.0 \times 10^{-3}$ & $6.0 \times 10^{-3}$ & 0.648 & 0.11  & $2.71 \times 10^{6}$ & - \\
\midrule 
(\textit{\citet{Shilpa2024}}) & & & & & & & \\
HHH & $4.0 \times 10^{-4}$ & $2.6 \times 10^{-4}$ & $2.6 \times 10^{-4}$ & 0.951 & $1.02$ & $1.44 \times 10^{6}$ & $1.25 \times 10^{6}$\\
MMM & $4.0 \times 10^{-4}$ & $1.6 \times 10^{-4}$ & $1.6 \times 10^{-4}$ & 0.896 & $1.04$ & $8.11 \times 10^{6}$ & $9.61 \times 10^{6}$\\
LLL & $4.0 \times 10^{-4}$ & $1.1 \times 10^{-4}$ & $1.1 \times 10^{-4}$ & 0.825 & $1.05$ & $3.05 \times 10^{7}$ & $1.25 \times 10^{7}$\\
\bottomrule
\end{tabular}
\caption{Viscous resistance (permeability) estimates obtained using the Kozeny-Carman model and comparison with measurements by \citet{Shilpa2024} for similar geometries.}
\label{tab:perm_com}
\end{table}

As a sensitivity test, numerical simulations were performed for the dense and transitional floating canopies subjected to ISWs using the exact viscous resistance values predicted by the KC model, and values multiplied by a factor of 5 lower and higher, i.e., corresponding to viscous resistance $5/K_{KC}$, $1/K_{KC}$ and $1/5K_{KC}$. These tests compared the predicted and measured horizontal velocity profiles at the wave trough for ISW of $a = 2$ cm under the long transitional and long dense canopies (A2L2T and A2L2D). Velocity profiles were compared at the streamwise location $x/h_1 = 10.5$ immediately downstream of the canopy, where experimental measurements were first available across the full water depth following the interaction between the ISW and the floating canopy. The difference was quantified using the $l_1$ norm, normalized by the nominal phase speed of the ISW, $c$, as defined in Eq.~\ref{eq:e_perm}:
\begin{equation}
    e = \frac{1}{N} \sum_{i=1}^N\left|\frac{u_{n, i}-u_{m, i}}{c}\right|,
    \label{eq:e_perm}
\end{equation}
where $u_n$ and $u_m$ are the numerical and measured horizontal velocity components, and $N$ is the total number of data points across the entire water depth.

\begin{table}
\centering
\begin{tabular}{l c c }
\toprule
Case & \qquad $1/K$ (m$^{-2}$) \qquad & \qquad $e$ \qquad \\
\midrule
\multirow{3}{*}{A2L2T}       & \quad $1.49 \times 10^5$  &  \quad 5.38 $\times 10^{-2}$\\
                              & \quad $2.97 \times 10^4$  &  \quad 6.13 $\times 10^{-2}$\\
                              & \quad $5.94 \times 10^3$  &  \quad 6.18 $\times 10^{-2}$\\
\midrule
\multirow{3}{*}{A2L2D}       & \quad $1.36 \times 10^7$  &  \quad 6.24 $\times 10^{-2}$\\
                              & \quad $2.71 \times 10^6$  &  \quad 6.49 $\times 10^{-2}$\\
                              & \quad $5.42 \times 10^5$  &  \quad 8.38 $\times 10^{-2}$\\
\bottomrule
\end{tabular}
\caption{Sensitivity tests evaluating the impact of the chosen viscous resistance on the integrated velocity error, $e$, from (\ref{eq:e_perm}).}
\label{tab:perm_slc}
\end{table}
 
\begin{figure}
    \centerline{\includegraphics[width = 1.0\textwidth]{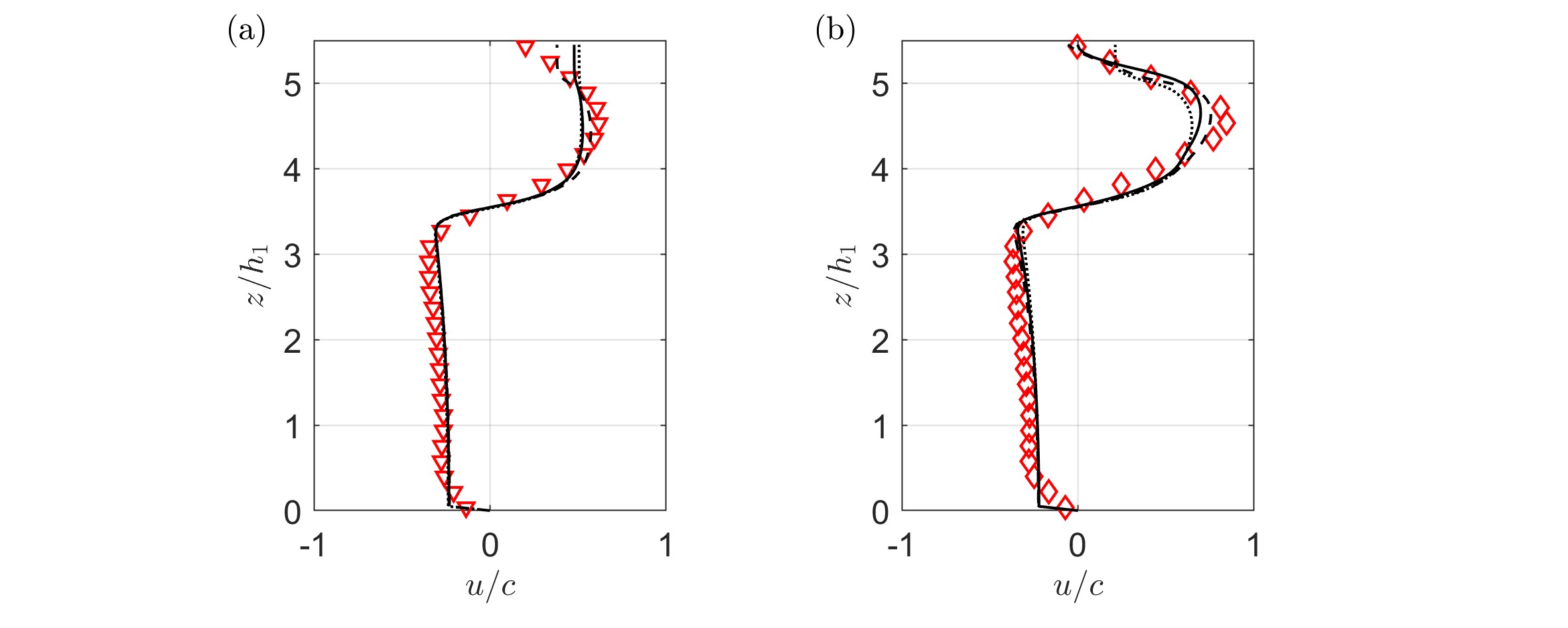}}
    \caption{Horizontal velocity profiles comparison between the experiments and the simulations of various viscous resistances. (a) A2L2T, (b) A2L2D. Solid curves $(1/K_{KC})$, dashed curves $(5/K_{KC})$, dotted curves $(1/5K_{KC})$.}
    \label{fig:k_inv_slc}
\end{figure}

Viscous resistance values scaled by a factor of $5$ from the KC model yielded better agreement with experimental results (Table~\ref{tab:perm_slc}). For both the transitional and dense canopies, this adjustment more accurately captured the acceleration within the upper-layer fluid and the deceleration downstream of the canopy structure, as demonstrated by the velocity profiles in Figure~\ref{fig:k_inv_slc}. Two factors may contribute to the consistently higher viscous resistance from the KC model. First, the unit cell used to compute the specific surface area (Figure~\ref{fig:sigma_unit}) neglects the contribution from the bottom stem surfaces. This exclusion may have a negligible impact on multi-layer canopy structures as those used in the experiments by \citet{Shilpa2024}. However, it becomes increasingly significant in single-layer structures like the floating canopies employed in the present study. In particular, the fully exposed lower interface of the canopy to the surrounding fluid likely enhances the impact of the bottom layer on overall resistance. Second, the KC model strictly applies only for the Darcy (creeping) flow regime, characterized by a low pore-scale Reynolds number, $Re_p < 1$. In the present study, an estimated pore-scale Reynolds number $Re_p = u d/ \nu > O(10)$ using the stem width, $d$, the horizontal velocity component within the upper layer fluid, $u$, and the kinematic viscosity of water, $\nu=10^{-6}$ m$^2$/s, places the flow in the transitional regime. The elevated $Re_p$ implies enhanced inertial effects and stronger resistance compared to the conditions assumed in the original KC formulation. Despite these limitations, the numerical simulations capture experimental trends reasonably well using viscous resistance values of $5/K_{KC}$.       

\section{Results}
\label{sec:res}

\subsection{Wave Amplitude Evolution}
\label{sec:isw_a_trans}

\begin{figure}
    \centerline{\includegraphics[width = 0.85\textwidth]{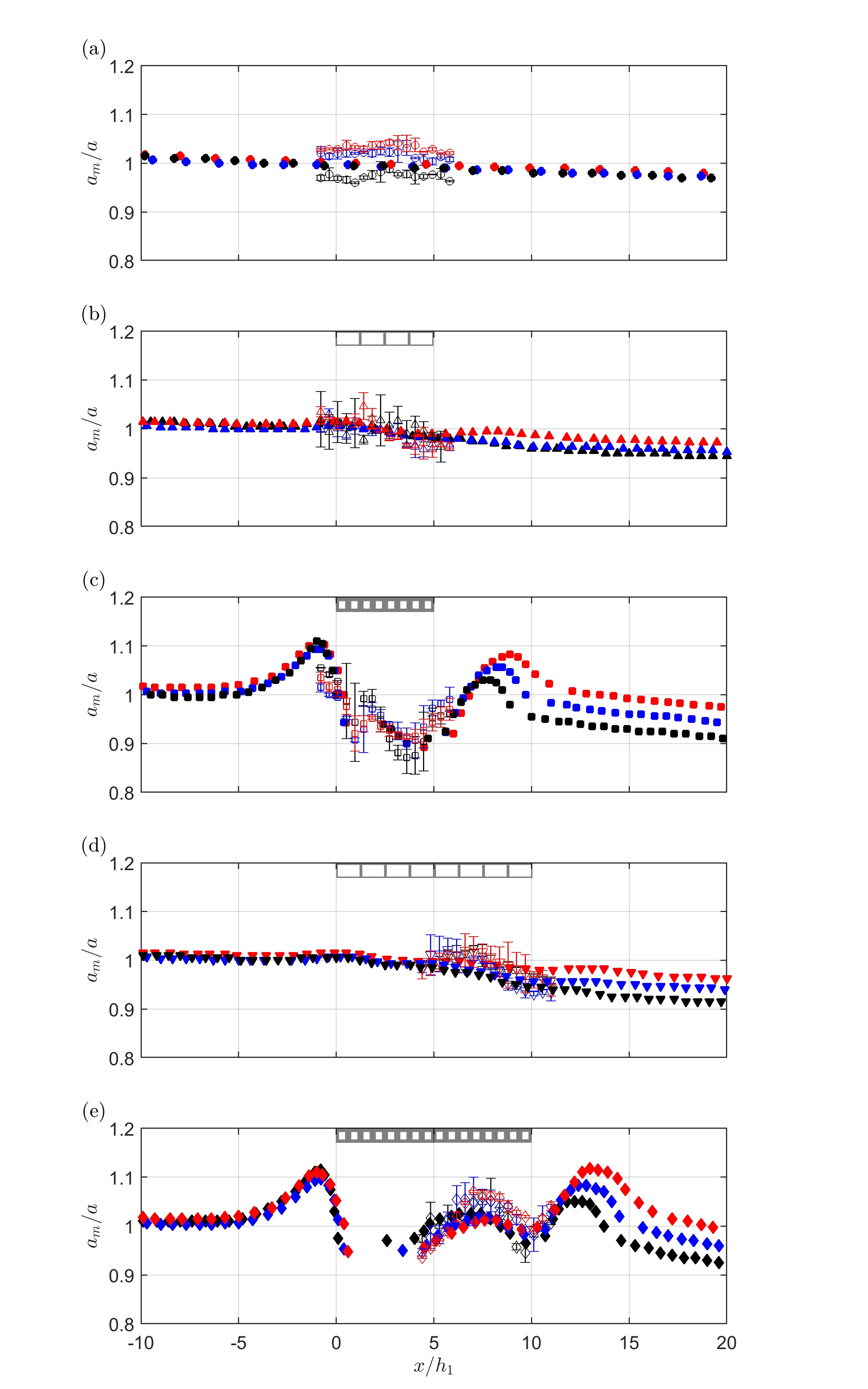}}
    \caption{Amplitude evolution in space for the (a) no canopy, (b) short transitional canopy (L1T), (c) short dense canopy (L1D), (d) long transitional (L2T), (e) long dense canopy (L2D) cases. Black, blue, and red markers correspond to ISW amplitudes $a = \{1, 1.5 ,2\}$ cm, respectively. }
    \label{fig:x_a}
\end{figure}

Figure~\ref{fig:x_a} presents the spatial evolution of ISW wave amplitudes under various canopy configurations. The pycnocline location $\zeta(t)$ is extracted from the PLIF visualization using the method outlined in Section 3.3 of \citet{Chu2025}, whereby a hyperbolic tangent profile is fitted to the observed vertical fluorescence intensity. The error bars represent the measured pycnocline thickness. In the numerical simulations, the pycnocline position is identified by locating the inflection point of the hyperbolic tangent density profile. To improve spatial precision beyond the grid resolution, the raw simulation data are interpolated to a refined vertical resolution of $dz = 1 \times 10^{-3}$ m prior to extracting the pycnocline profile. Experimental data are represented by hollow markers as listed in Table~\ref{tab:case_exp}, while the simulation results are represented by the corresponding solid markers. 

In the absence of a floating canopy (Figure~\ref{fig:x_a}(a)), wave amplitudes remain nearly constant throughout the observation window in the experiments and simulations. For the transitional canopy cases (Figure~\ref{fig:x_a}(b,d)), the wave amplitudes show dissipation of around 5$\sim$10\%, depending on the canopy length.  There is a modest increase in wave amplitude at the leading edge of the canopy. 

For the dense canopy cases (Figure~\ref{fig:x_a}(c,e)), pronounced non-linear interactions are observed. The numerical simulations show an initial increase in wave amplitude upstream of the canopy, beginning at $x/h_1\approx-5$.  This is followed by a sharp decay around $x/h_1\approx0$. This decay is observed in both numerical simulations and experimental measurements for the short dense canopy. Afterwards, a complex wave transformation process occurs over $0<x/h_1<5$. In the second half of the long dense canopy ($5<x/h_1<8$), the wave amplitudes show a mild recovery and plateau. Downstream of the dense canopy, both the experiments and simulations show evidence of amplitude enhancement. The experimental PLIF measurements do not extend significantly beyond the canopies. However, the numerical simulations predict that this amplitude enhancement is sustained over a distance of $\approx 5h_1$ beyond the canopy.  These complex transition processes at the leading and trailing edges of the dense canopies are described in greater detail in the following subsections.

Experimental results for the floating canopies, where available, generally show good agreement with the numerical predictions. Moreover, ISW amplitude does not appear to play a major role in dictating wave evolution underneath the canopies; the normalized amplitude profiles show reasonable collapse across all cases. Downstream of the canopy, however, the numerical simulations do show stronger wave amplitude dependence.

The trends observed in Figure~\ref{fig:x_a} indicate that the canopy density can play a significant role in shaping wave transformation. Although the present study only considers two canopy porosities, the results indicate that the geometric thresholds identified in prior work to delineate sparse and dense canopy behavior (i.e., a strong shear layer only develops at the interface for $\lambda_f \gg 0.1$) remain appropriate for floating canopy - ISW interactions.

\subsection{Leading Edge Shoaling}
\label{sec:isw_ld_shoal}

\begin{figure}
    \centerline{\includegraphics[width = 1.0\textwidth]{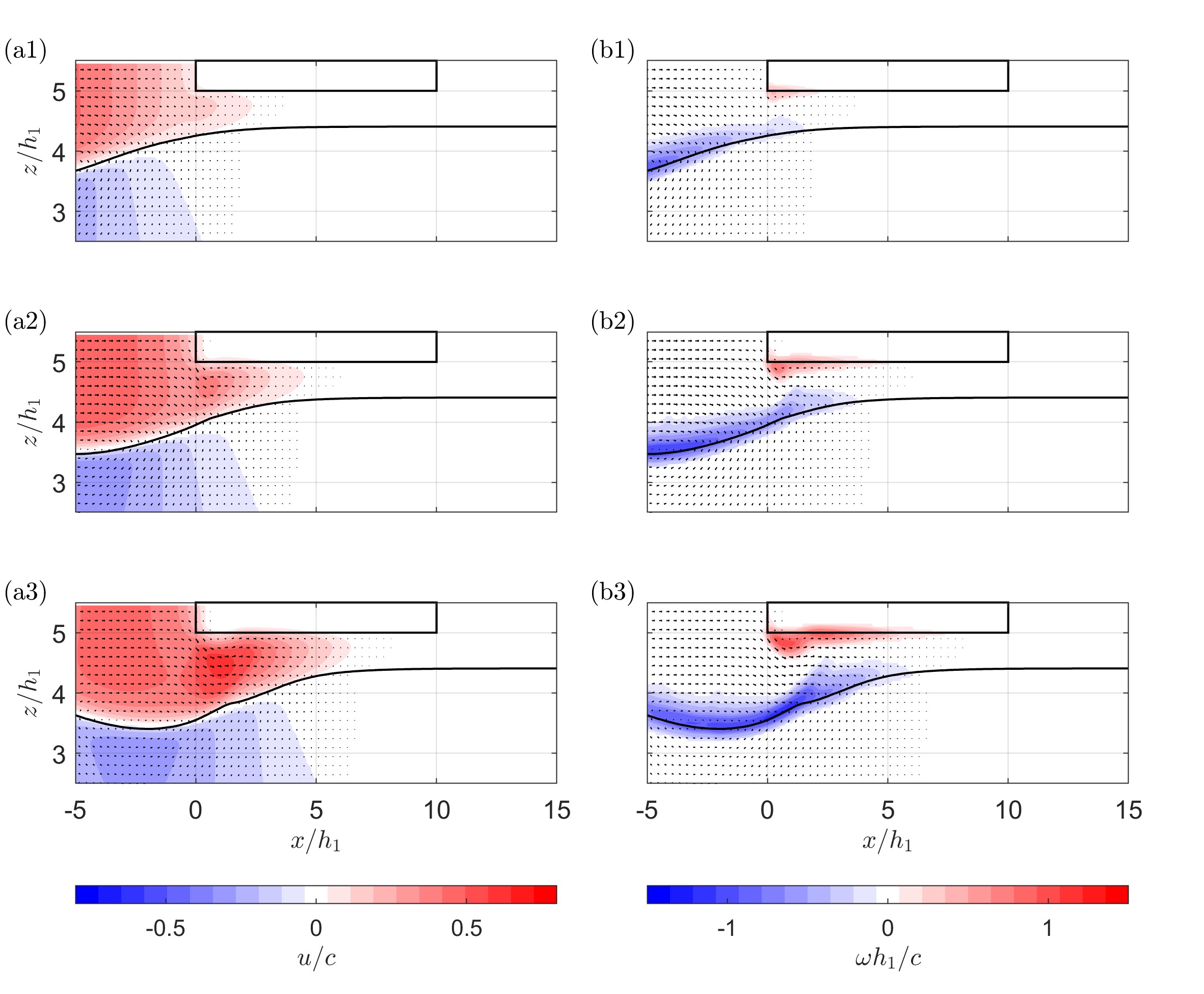}}
    \caption{Simulation results for case A2L2D illustrate the leading edge shoaling process. Panels (a1–a3) show consecutive normalized horizontal velocity contours in time, and panels (b1–b3) show consecutive normalized vorticity contours in time. The solid black lines denote the pycnocline and floating canopy.}
    \label{fig:ld_shoal}
\end{figure}

As the ISW encounters the front edge of the dense canopy, the upper layer flow is obstructed and diverted below the structure. This obstruction leads to a localized amplitude growth starting from $x/h_1 \approx -5$, as shown in Figure~\ref{fig:x_a}(c,e). Meanwhile, a positive vorticity packet begins to develop at the bottom edge of the canopy, as observed in Figure~\ref{fig:ld_shoal}(b1-b3). This vorticity growth is driven by enhanced shear at the canopy interface, induced by the diversion of high-momentum upper layer fluid underneath the canopy as the ISW trough approaches. This shear and the intrinsic shear generated at the ISW pycnocline give rise to a counter-rotating vortex pair. The resulting vortex-induced jet accelerates the leading side of the ISW, as illustrated in Figure~\ref{fig:ld_shoal}(a3). This process accounts for the rapid amplitude drop around $x/h_1 \approx 0$ observed in the dense canopy cases in Figure~\ref{fig:x_a}(c,e). 

In contrast, for the transitional canopy cases, the upper layer fluid is able to penetrate the canopy structure more easily. This yields a milder horizontal velocity deficit and reduced shear. As a result, the amplitude growth caused by flow obstruction as well as the subsequent amplitude reduction induced by the vortex-driven jet is much less pronounced. This trend is reflected in Figures~\ref{fig:x_a}(b,d), where the wave amplitude remains relatively stable.

\subsection{Soliton Adjustment}
\label{sec:isw_sol_adj}

\begin{figure}
    \centerline{\includegraphics[width = 1.0\textwidth]{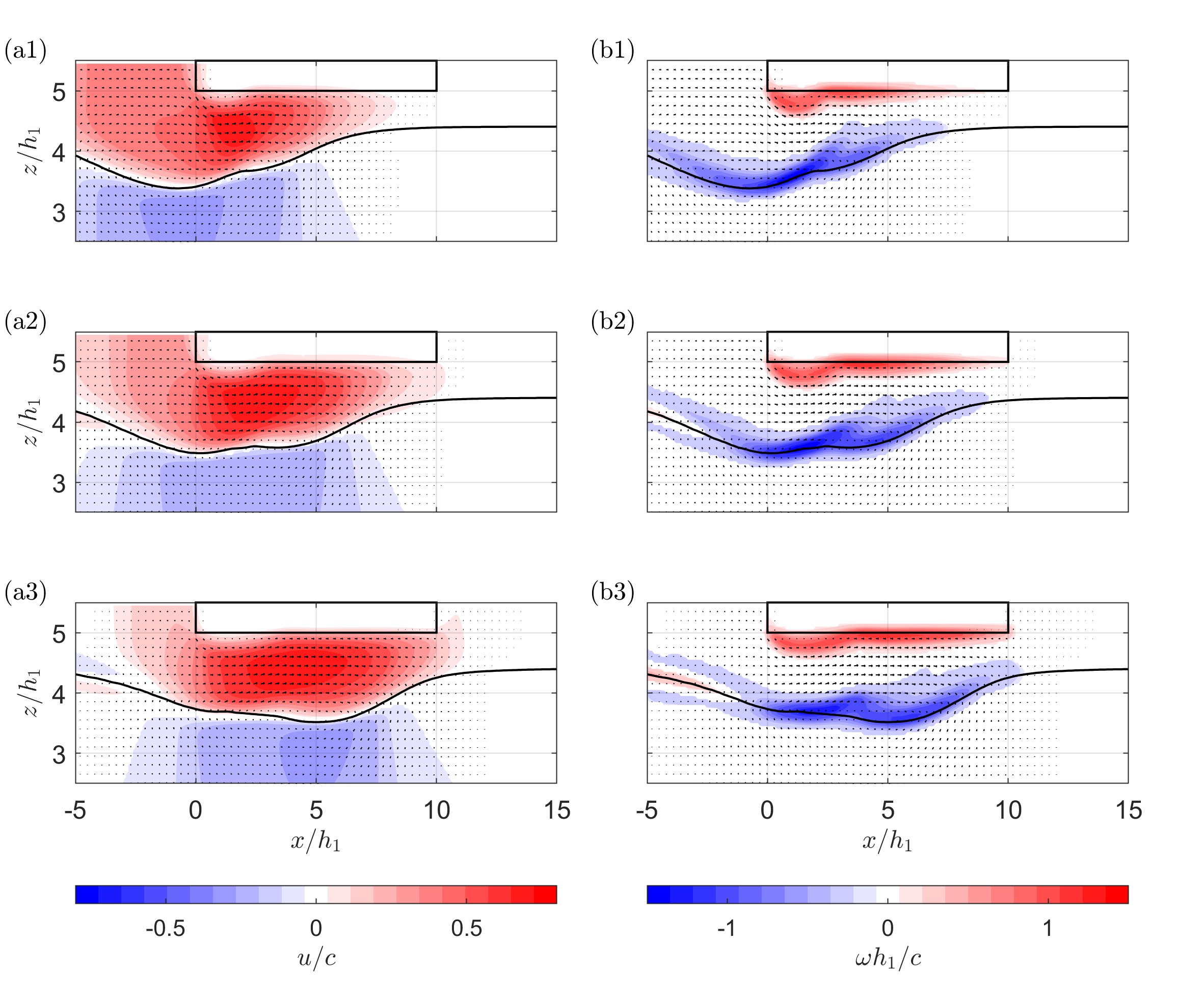}}
    \caption{Simulated results of case A2L2D illustrate the soliton adjustment process. Panels (a1–a3) show consecutive normalized horizontal velocity contours in time and panels (b1–b3) show consecutive normalized vorticity contours in time. The solid black lines denote the pycnocline and floating canopy.}
    \label{fig:sol_adjust}
\end{figure}

\begin{figure}
    \centerline{\includegraphics[width = 1.0\textwidth]{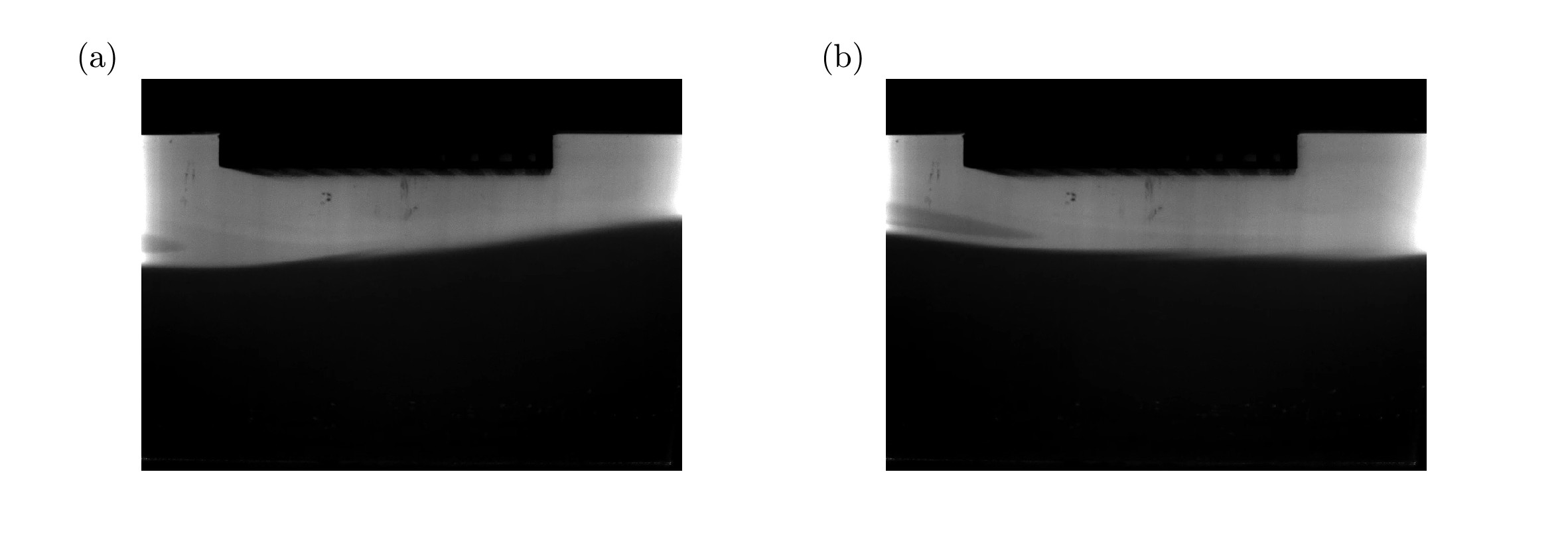}}
    \caption{PLIF visualization of soliton adjustment for case A2L1D.}
    \label{fig:plif_sol_adjust}
\end{figure}

As the ISW propagates further into the canopy, the vortex-induced jet mechanism described in Section~\ref{sec:isw_ld_shoal} continues to shift upper layer fluid volume from the trailing edge of the wave towards its leading edge. The volume shift adjusts the relative position of the wave trough forward in space (Figure~\ref{fig:sol_adjust} (a1-a3)), creating a double-trough pycnocline structure underneath the canopy. The soliton adjustment process occurs over $0 < x/h_1 < 5$, as shown by the simulation and experimental results in Figures~\ref{fig:x_a}(c,e). 

The forward adjustment of the soliton is directly captured in the PLIF visualization of the A2L1D configuration (Figure~\ref{fig:plif_sol_adjust}). At a earlier time step, the wave trough remains positioned near the trailing edge of the ISW, as shown in Figure~\ref{fig:plif_sol_adjust}(a), consistent with the simulation result in Figure~\ref{fig:sol_adjust}(a1). At a later time step, the vortex-driven jet shifts the trough forward toward the leading edge of the ISW (Figure~\ref{fig:plif_sol_adjust}(b)), matching the adjustment process captured in the simulation at Figure~\ref{fig:sol_adjust}(a3).

\subsection{Phase Speed Adjustment}
\label{sec:isw_c_trans}

\begin{figure}
    \centerline{\includegraphics[width = 1.0\textwidth]{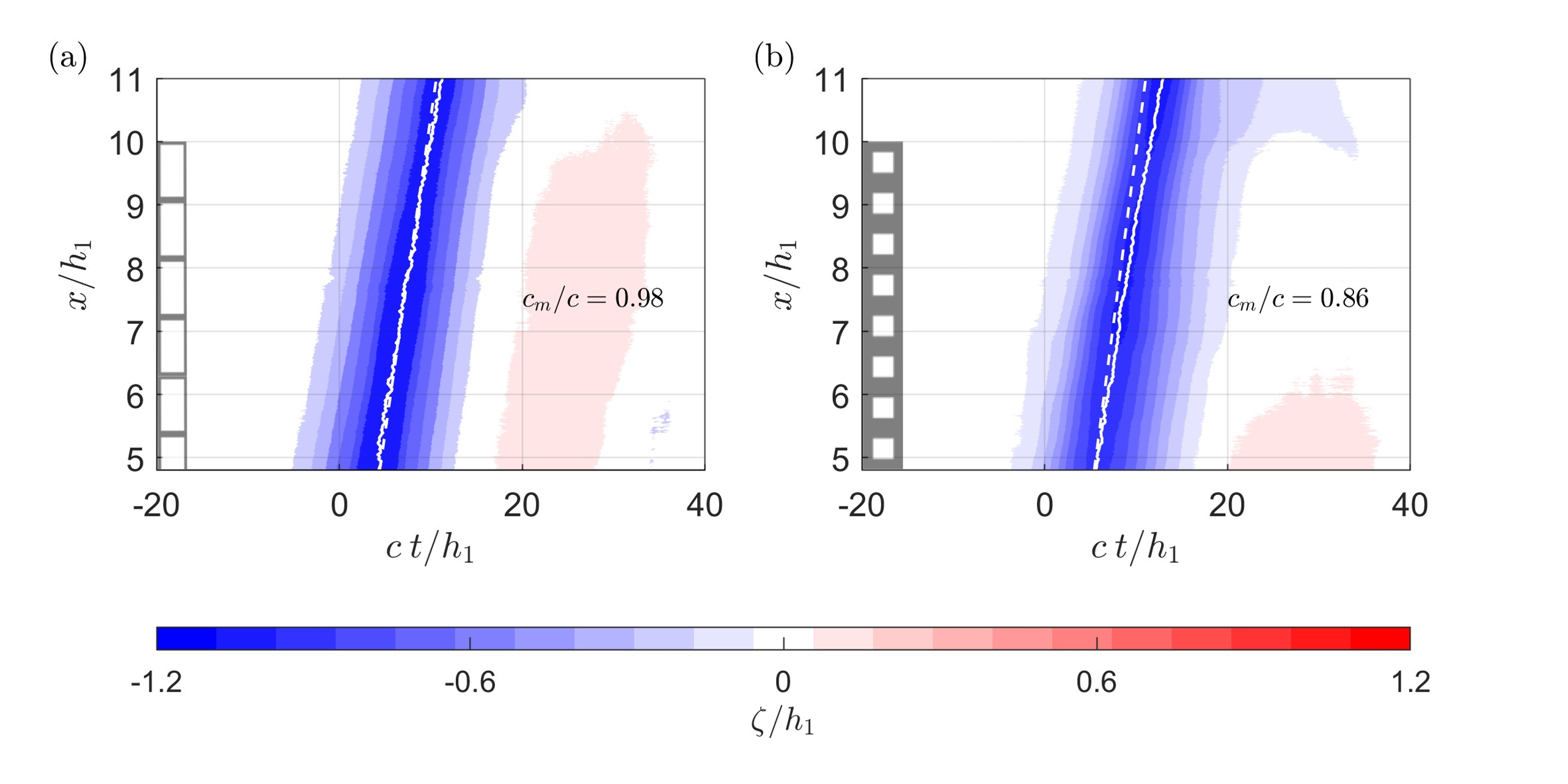}}
    \caption{Hovmöller diagrams showing the evolution of normalized wave profiles (i.e., $\zeta(x,t)/h_1$) obtained from PLIF measurements for the (a) A2L2T configuration and (b) A2L2D configurations. Solid curves indicate the tracked wave trough locations from PLIF data, while dashed lines represent phase speed predictions from the eKdV solution based on nominal layer depths and wave amplitudes.}
    \label{fig:hovmoller}
\end{figure}

\begin{figure}
    \centerline{\includegraphics[width = 1.0\textwidth]{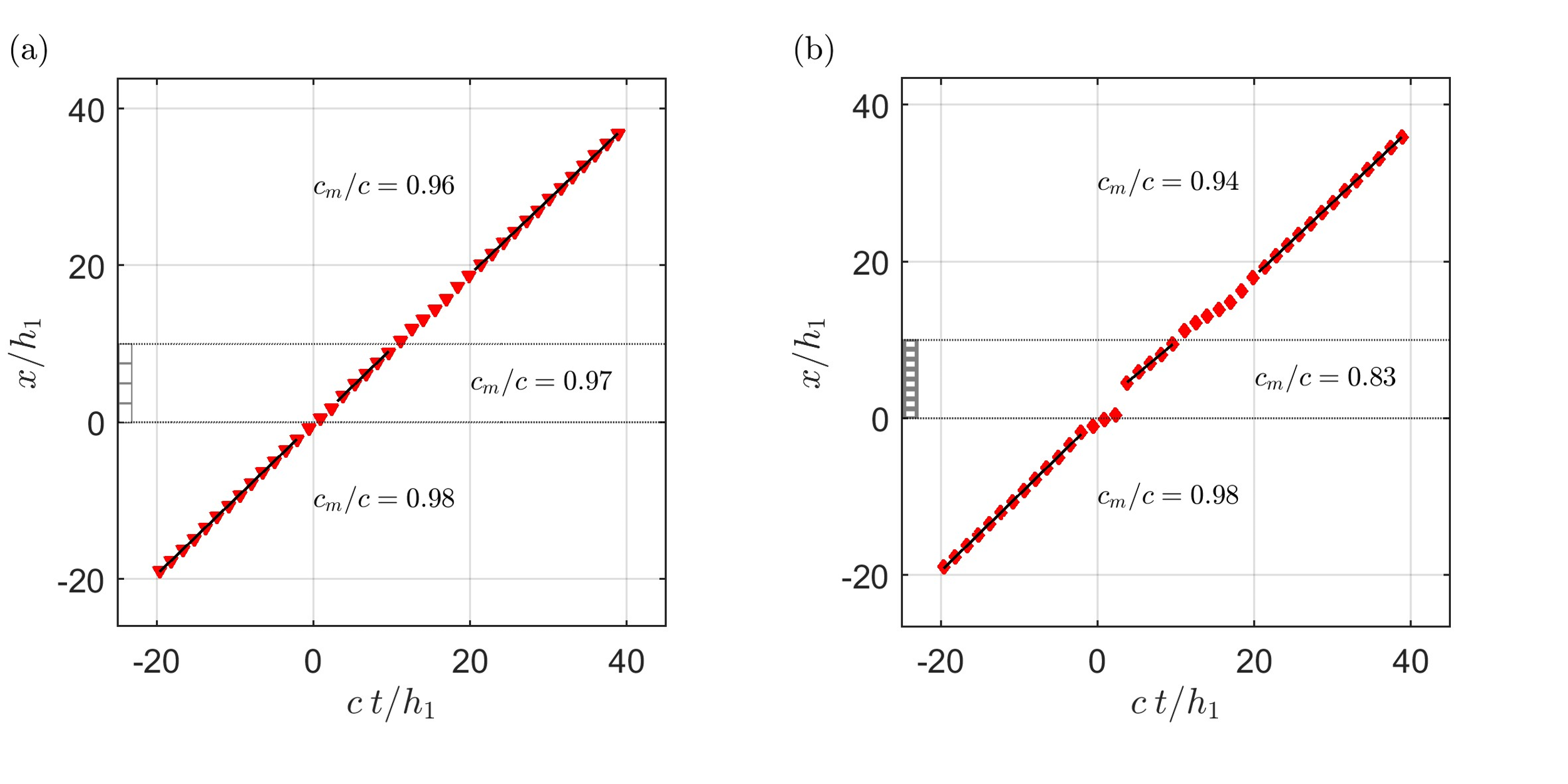}}
    \caption{Simulation results tracking ISW trough location for the (a) A2L2T and (b) A2L2D cases.}
    \label{fig:phase_speed}
\end{figure}

Immediately following the soliton adjustment process at the leading side of the canopy, amplitude recovery is observed in the second half of the long, dense (L2D) canopy cases (Figures~\ref{fig:x_a}(e)). This recovery is triggered by a phase speed adjustment. For the transitional canopy cases, the trough trajectory closely follows the prediction of the eKdV solution (Figures~\ref{fig:hovmoller}(a)). Upstream of the L2T canopy, the measured phase speed, $c_m$, in the simulation is 2\% slower than the theoretical value  (Figures~\ref{fig:phase_speed}(a)). Below and downstream of the L2T canopy, the phase speed remains largely unaffected, with only a marginal 1\% reduction because of the dissipation induced by the L2T canopy. 

In contrast, wave propagation underneath the L2D canopy shows significant deviation from the eKdV prediction (Figures~\ref{fig:hovmoller}(b)). Upstream of the L2D canopy, the measured phase speed matches that of the L2T case.  However, immediately after the soliton adjustment process described in the previous section, the trough trajectory stabilizes into an apparent steady state with a roughly 15\% reduction in phase speed relative to the upstream condition (Figures~\ref{fig:phase_speed}(b)). These observations are consistent across the experiments and simulations. Downstream of the L2D canopy, simulations show that the phase speed partially recovers, ending up with a residual 4\% reduction compared to the upstream value. A similar wave amplitude recovery process can be observed in the numerical study by \citet{Terletska2024} evaluating the interaction of an ISW with a floating ice sheet. Since wavelengths of the generated ISWs exceeds canopy length of the long canopy, the quasi-steady propagation lasts less than $3h_1$ after phase speed adjustment. This is immediately followed by the edge effect downstream of the canopy (Figures~\ref{fig:x_a}(e), $8 < x/h_1 <12 $). To isolate the phase speed adjustment from these transient edge effects, additional simulations on a dense canopy of $l_c = 60h_1$ are presented in Appendix \ref{appA}.      

\subsection{Trailing Edge Steepening}
\label{sec:isw_vort_stp}

\begin{figure}
    \centerline{\includegraphics[width = 1.0\textwidth]{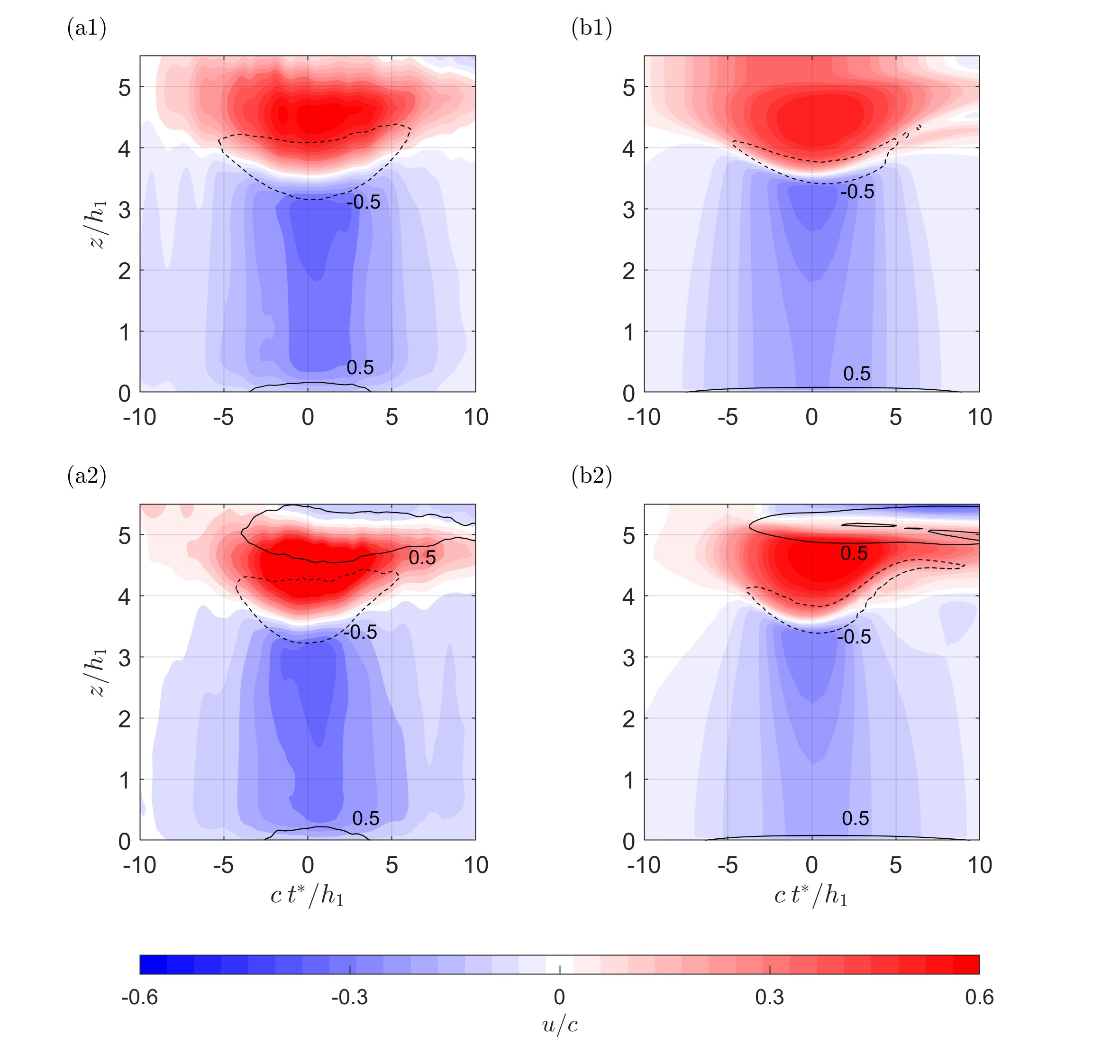}}
    \caption{Contour plots showing normalized profiles of horizontal velocity at $x/h_1 = 10.5$ as a function of time. Upper panels show the A2L2T configuration and lower panels show the A2L2D configuration. Panels (a1, a2) present experimental PIV measurements and (b1, b2) display the corresponding simulations. Solid and dashed contours mark regions of normalized vorticity at $\omega h_1/c = 0.5$ and $-0.5$, respectively.}
    \label{fig:u_hovmoller}
\end{figure}

Figure~\ref{fig:x_a}(c,e) show a significant increase in ISW amplitude downstream of the dense canopies. This wave profile steepening is the result of wave dynamics coupling with vortex dynamics at the trailing edge of the canopy. Figure~\ref{fig:u_hovmoller} shows normalized profiles for the horizontal velocity downstream of the long floating canopy. The temporal axis is shifted by subtracting the arrival time of the ISW trough at $x/h_1 = 10.5$, whereby $t^* = 0$ represents the trough arrival time at this location. For the transitional canopy configuration  (case A2L2T) shown in the upper panels, the upper layer fluid is able to penetrate the floating canopy. This results in minimal shear at the canopy interface and limits flow separation downstream of the canopy. Only a minor flow reversal is observed near the end of the interaction at $c t^*/h_1 \approx 9$. Therefore, only the negative vorticity associated with the intrinsic shear of the ISW, resulting from the velocity gradient at the pycnocline, is observed in Figure~\ref{fig:u_hovmoller}(a1,b1). 
For the dense canopy configuration (case A2L2D) shown in the lower panels, significant flow separation develops downstream of the canopy after $c t^*/h_1 \approx -2.5$, and this separation lasts beyond $c t^*/h_1 = 10$. This enhanced separation, due to the blockage imposed by the dense canopy, gives rise to an additional positive vorticity pocket around $z/h_1 \approx 5$. 

\begin{figure}
    \centerline{\includegraphics[width = 1.0\textwidth]{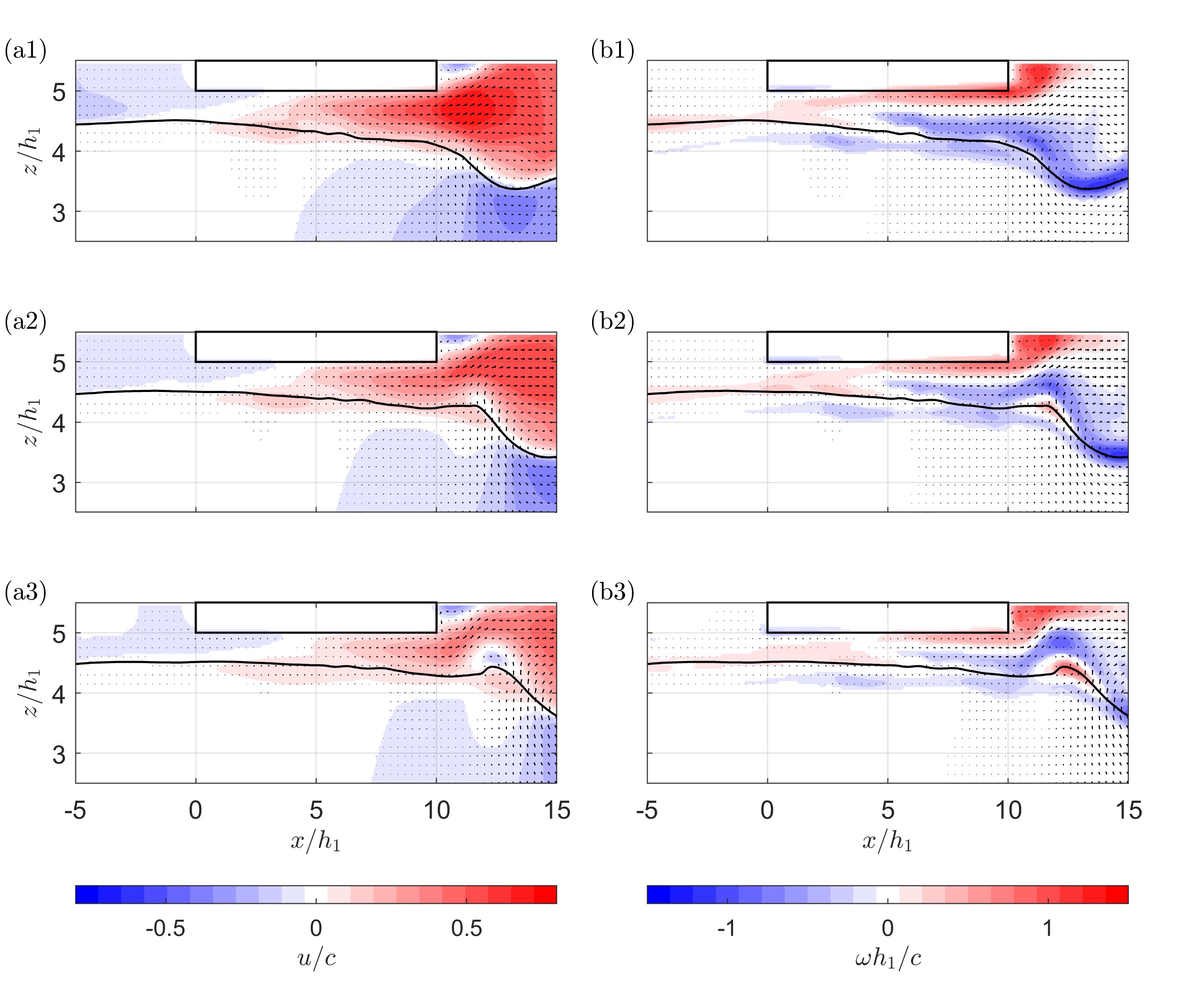}}
    \caption{Simulated results of case A2L2D illustrate the vortex-driven steepening process. (a1–a3) Consecutive normalized horizontal velocity contours in time, (b1–b3) consecutive normalized vorticity contours in time. The solid black lines denote the pycnocline and floating canopy.}
    \label{fig:vor_stp}
\end{figure}

The flow separation for the A2L2D configuration, captured in Figure~\ref{fig:u_hovmoller}(a2,b2), is evident immediately downstream of the canopy in the simulation results shown in Figure~\ref{fig:vor_stp}. The resulting positive vortex from the flow separation pairs with the the negative vortex at the pycnocline, inducing a strong jet. 
The jet transfers volume and momentum within the upper layer fluid forward from the trailing side of the ISW to the wave trough, leading to wave steepening. The vortex-induced jet is sufficient to trigger a secondary flow reversal at the pycnocline ($x/h_1 \approx 12$ in Figure~\ref{fig:vor_stp}) to locally maintain stability, where the buoyancy gradient prevents the upward penetration of the denser salt solution. This secondary flow reversal leads to the formation of a small positive vortex at the pycnocline.

Similar vortex-induced jet formation and trailing edge steepening mechanisms have been reported in the numerical study by \citet{Zhang2022} on internal solitary wave evolution beneath an ice keel. In their simulations, an ISW with an initial amplitude of $|a|=12$ m under the depth ratio of $(h_1,\, h_2) = (20,\, 80)$ m encountered a fixed ice keel with a height equal to the upper-layer depth $(h_k = h_1)$. As the ISW trough passed over the tip of the ice keel, an intensified flow was observed directly downstream. The wave amplitude increased by approximately 50\% from the initial value to $|a| \approx 18$ m downstream of the ice keel. Compared to the present study, additional overturning was observed downstream of the ice keel. This is to be expected for a solid structure with a larger obstruction relative to the upper-layer depth as compared to the floating canopies considered in this paper. 

\subsection{Wave Energy}
\label{sec:isw_e_trans}
The trends observed in the preceding sections can also be evaluated in terms of wave energy budgets. The total energy $(E_t)$, defined as the sum of instantaneous kinetic energy $(E_k)$ and potential energy $(E_a)$, is calculated as follows \citep{Zhang2022, Forgia2021}:
\begin{equation}
    E_{t}(t) = E_k(t) + E_a(t),
    \label{eq:Et}
\end{equation}
where $E_k$ and $E_a$ are respectively expressed as: 
\begin{equation}
    E_k\left(t\right)=\frac{1}{2} \iint_D \rho\left(x, z, t\right)\left[u\left(x, z, t\right)^2+w\left(x, z, t\right)^2\right] d x d z,
    \label{eq:Ek}
\end{equation}
and
\begin{equation}
    E_a\left(t\right)=g \iint_D\left[\rho\left(x, z, t\right)-\rho^*\left(x, z, t\right)\right] z d x d z .
    \label{eq:Ea}
\end{equation}
In these expressions, $t$ is the time, $u(x,z,t_i)$ and $w(x,z,t_i)$ are the horizontal and vertical velocity components, $\rho(x,z,t_i)$ is the density distribution, and $\rho^*(x,z,t_i)$ is the density distribution for the stably stratified background condition. Thus, $E_a$ quantifies the potential energy required to displace a stably stratified fluid into the perturbed state, such as the ISW wave profile. The spatial integration domain, $D$, covers the full water depth $h_1+h_2$ in the vertical direction, and $30h_1$ in the horizontal direction centered at the wave trough, which encompasses a region slightly larger than the ISW wavelength. Since the energy terms are evaluated over a 2D cross section, their units are expressed in J/m.

\begin{figure}
    \centerline{\includegraphics[width = 1.0\textwidth]{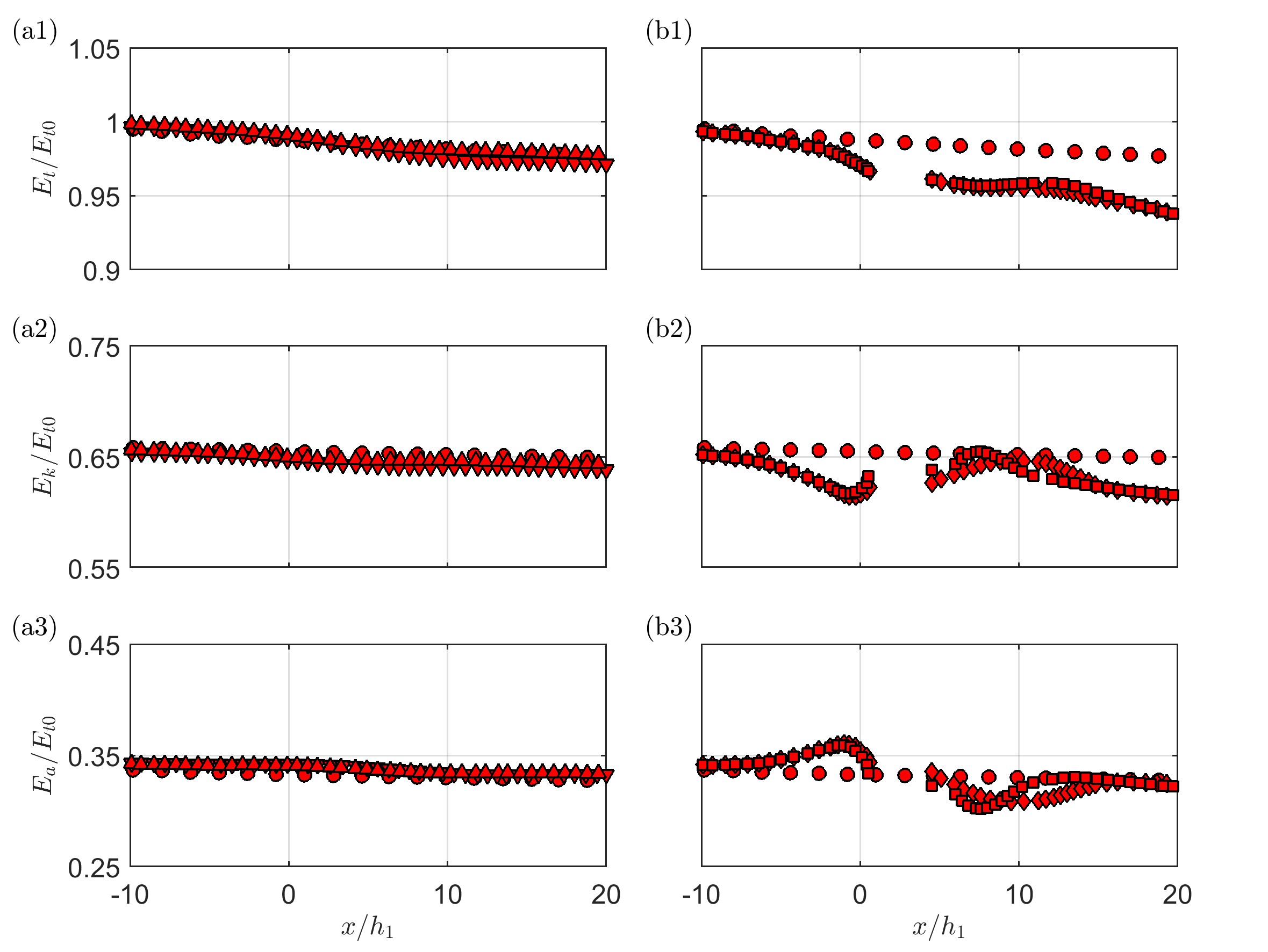}}
    \caption{Spatial variation of energy by initial total energy. Panels (a1,b1) show total energy, (a2,b2) show kinetic energy, and (a3,b3) available potential energy. Panels (a1)-(a3) show the transitional canopy cases (A2L1T, A2L2T) with the red circles representing the corresponding no-canopy case. Panels (b1)-(b3) show the dense canopy cases (A2L1D, A2L2D) with the red circle markers again showing the no-canopy reference.}
    \label{fig:x_e}
\end{figure}

Figure~\ref{fig:x_e} shows the spatial evolution of different energy components normalized by the initial total energy. In the absence of a canopy, the total energy decreases by approximately 3\% over a distance of $30h_1$ (Figure~\ref{fig:x_e} (a1, b1)), primarily due to the dissipation inherent in the simulations. For the transitional canopy cases (L1T and L2T), the energy profiles closely follow the trend observed in the no-canopy scenario, indicating minimal additional energy dissipation (Figure~\ref{fig:x_e}(a1–a3)). In contrast, the dense canopy cases (L1D and L2D) exhibit a further 3–4\% reduction in total energy relative to the no-canopy condition (Figure~\ref{fig:x_e}(b1)). Energy exchange between $E_k$ and $E_a$ is observed throughout these processes (Figure~\ref{fig:x_e}(b2,b3)). During the leading edge shoaling ($-5<x/h_1<2$), part of $E_k$ is converted into $E_a$ in the form of local amplitude growth. As the phase speed adjusts beneath the second half of the longer L2D canopy ($5<x/h_1<8$), a portion of $E_a$ is transferred back into $E_k$. Finally, a small amount of $E_k$ is restored to $E_a$ due to trailing edge steepening ($8<x/h_1<10$).

\begin{figure}
    \centerline{\includegraphics[width = 1.0\textwidth]{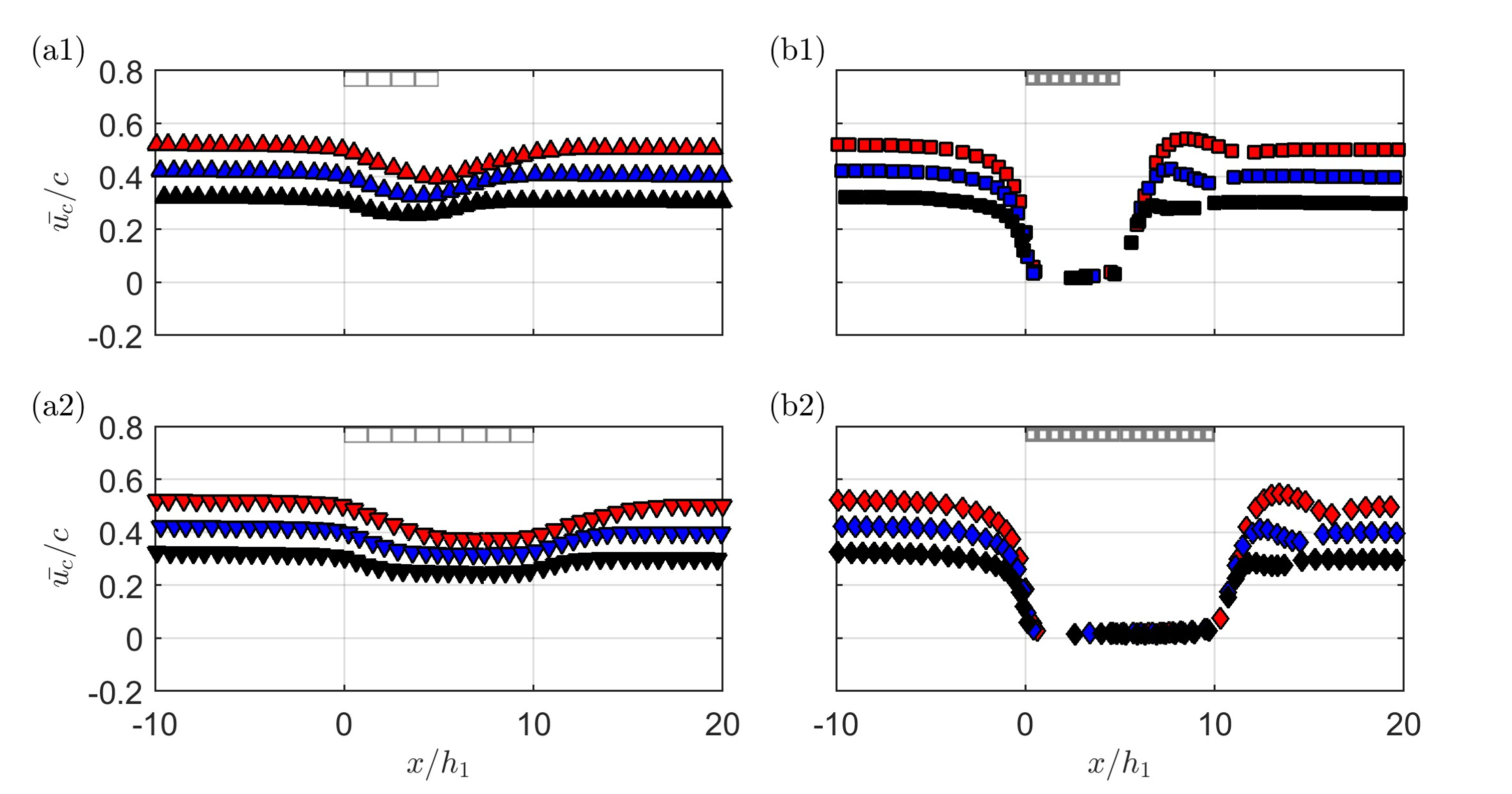}}
    \caption{Variation of mean horizontal velocities within the canopy during the passage of the wave trough for the cases corresponding to canopy configurations (a1) L1T, (b1) L1D, (a2) L2T, and (b2) L2D. Black, blue and red markers correspond to nominal ISW amplitudes of $a = \{1, 1.5, 2\}$ cm, respectively.}
    \label{fig:x_uc}
\end{figure}

Total energy reduction in the dense canopy cases occurs  primarily at leading edge shoaling ($-5<x/h_1<2$). Canopy length does not play a significant role for total energy dissipation, as suggested by the near overlap of the A2L1D and A2L2D dissipation curves (Figure~\ref{fig:x_e}(b1)). The weak canopy length dependence results from the rapid flow adjustment under dense canopies. As shown in Figure~\ref{fig:x_uc}(b1,b2), there is a substantial reduction in the mean in-canopy horizontal velocity during leading edge shoaling, which limits additional dissipation afterwards. In figure~\ref{fig:x_uc}, $\bar{u_c}$ denotes the mean horizontal velocity within the canopy height ($z/h_1 > 5$) evaluated at the wave trough. In contrast, under transitional canopies (Figure~\ref{fig:x_uc} (a1, a2)), $\bar{u_c}$ adjusts over a longer distance, exceeding $x/h_1 = 5$. The difference in the development length of $\bar{u_c}$ between the transitional and dense canopies is also consistent with prior observations \citep{rominger2011flow}, which indicate that flow adjustment around canopies would occur over the length scale $\sim h_c / \lambda_f$.  For the dense canopy with $\lambda_f \approx O(10^0)$, this adjustment would take place over a distance of $h_c \approx h_1/2$.  For the transitional canopy with $\lambda_f \approx O(10^{-1})$, this adjustment would take place over a distance of $h_c/0.1 \approx 5h_1$. These scales are broadly consistent with the observations in Figure~\ref{fig:x_uc}.


\begin{figure}
    \centerline{\includegraphics[width = 1.0\textwidth]{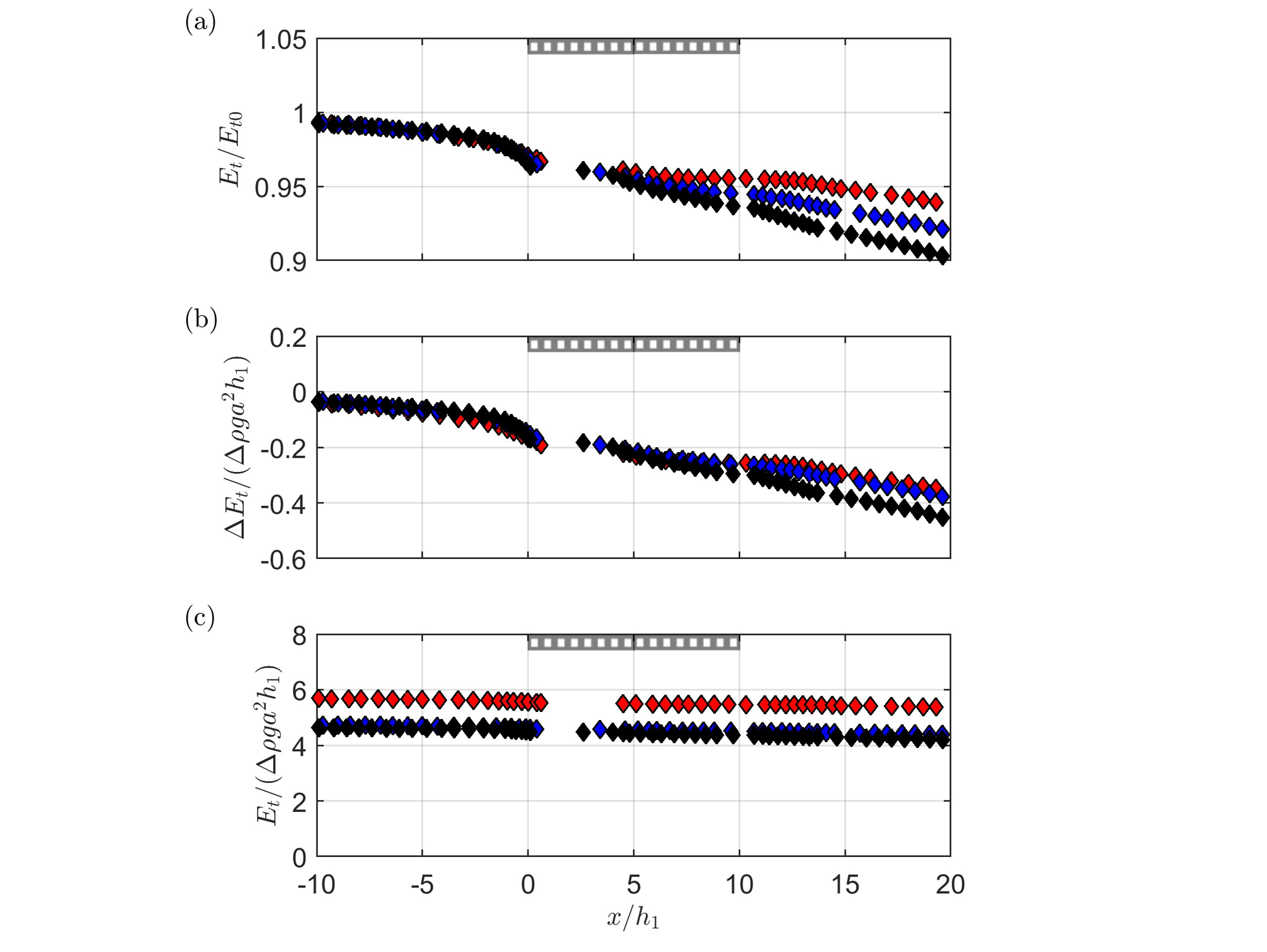}}
    \caption{Spatial variation of (a) total energy normalized by initial value, (b) scaled total energy difference, normalized by $a^2$, scaling and (c) scaled total energy normalized by $a^2$ scaling. All cases correspond to the long dense (L2D) canopy. Black, blue, and red markers correspond to ISW amplitudes $a = \{1, 1.5 ,2\}$ cm.}
    \label{fig:x_e_L2D_scaled}
\end{figure}

Under the same L2D configuration, the ISW of a small wave amplitude is subject to stronger total energy dissipation in percentage terms, as shown in Figure~\ref{fig:x_e_L2D_scaled}(a). Spatial variations of the scaled total energy difference, $\Delta E_t/(\Delta \rho g a^2 h_1)$, and scaled total energy, $E_t/(\Delta \rho g a^2 h_1)$, are presented in Figure~\ref{fig:x_e_L2D_scaled}(b,c). After normalized by $(a^2)$, curves for the scaled total energy \textit{difference} relative to the initial value nearly collapse together, indicating that the dissipation processes depend on amplitude squared. However, the scaled total energy for the large-amplitude case with $|a| = 2$ cm is slightly higher than the remaining cases. As discussed by \citet{Lamb2009}, the total energy of an ISW does not follow a fixed polynomial scaling, but a nonlinear function of the layer depth ratio and wave amplitude. For small-amplitude waves, the ISW profiles can be approximated by a weakly nonlinear model, yielding an approximate $a^2$ scaling of the total energy. However, as the wave amplitude approaches the theoretical limit for a given layer depth ratio, the wavelength and wave profile expand rapidly \citep{Helfrich2006}, causing both $E_a$ and $E_k$ to scale with higher orders of $a$ due to the increased volume flux of high-momentum fluid. Thus, while the energy dissipation scales approximately with $a^2$, the total energy may scale at a higher order, particularly for large-amplitude waves near the theoretical eKdV limit. As a result, the canopy-induced dissipation accounts for a relatively larger fraction of the total energy for small-amplitude waves. This is confirmed by the additional long-canopy simulations presented in Appendix~\ref{appA}, which show the highest normalized amplitude reductions for ISWs with amplitude $|a|=1$ cm.            

The ISW of $|a| = 1$ cm experiences around 7\% of total energy dissipation with the L2D configuration as illustrated in Figure~\ref{fig:x_e_L2D_scaled}(a). In comparison, approximately 10\% of the total energy was dissipated, and less than 3\% was reflected for the ISW of similar layer depth ratio and wave amplitude interacting with a solid ice keel \citep{Zhang2022}. In this prior study, reflection only became the dominant mechanism over dissipation when the obstruction thickness exceeded the upper-layer depth. As such, the observed dissipation for the porous obstructions considered here is consistent with expectations, and the reflected energy component is expected to be less important. 
  
\section{Conclusion}
\label{sec:conclusion}

The interaction between internal solitary waves (ISWs) and floating canopy structures was studied both numerically and experimentally. Laboratory measurements were used to validate complementary Reynolds-Averaged Navier-Stokes (RANS) simulations. The hydraulic resistance of the canopy in numerical simulations was modeled using the established Kozeny-Carman (KC) model for porous media. This led to reasonable agreement in the observed wave transformation processes and velocity profiles across the simulations and experiments. However,  empirical tuning of the porous viscous resistance in the KC model was necessary to achieve this agreement, indicating that additional work is needed to accurately characterize canopy resistance. 

The results indicated that the less dense (transitional) canopy exerted minimal impact on ISW propagation, producing limited amplitude reduction and negligible additional energy dissipation relative to the no-canopy condition. In contrast, the enhanced shear at the canopy interface generated by the dense canopy interacted with the opposing shear along the pycnocline, giving rise to complex nonlinear dynamics. During the leading edge shoaling stage, canopy obstruction resulted in volume accumulation and thus local amplitude growth upstream of the canopy. As the shear layer developed at the canopy interface, interactions with the opposing shear along the pycnocline generated a vortex-induced jet which accelerated the upper-layer fluid in the leading half of the ISW. This caused a rapid amplitude drop at the leading edge of the canopy. As the ISW traveled further beneath the canopy structure, it settled into a quasi-steady state with a reduced phase speed. This phase speed adjustment promoted secondary amplitude growth under the second half of the long dense canopy. As the wave trough passed over the canopy's trailing edge, flow separation was observed downstream of the structure. A vortex-induced jet mechanism transferred volume and momentum forward from the trailing side of the ISW into the wave trough, resulting in wave steepening. 
    
Wave energy analysis revealed that the dense canopy with porosity $n=0.648$ and canopy height $h_c = 0.5h_1$ produced an additional total energy dissipation of 3-7\%, depending on the wave amplitude, compared to the no-canopy case. Owing to the nonlinear nature of ISWs, smaller-amplitude waves experienced greater dissipation in percentage terms compared to larger-amplitude waves. While the dissipated energy scaled approximately with $O(a^2)$, the total energy did not follow a fixed amplitude scaling. For weakly nonlinear, small-amplitude ISWs, the total energy exhibited an $O(a^2)$ dependence. However, increased nonlinearity in larger-amplitude waves introduced higher-order scaling, associated with an expanded wave profile and broader high-momentum flux. In contrast to its dependence on wave amplitude, the total energy dissipation showed a weak dependence on the canopy length for the dense canopy cases. The total energy dissipation was associated with the drag produced by the canopy structure, where the mean horizontal velocity component within the canopy depth vanished during the early flow adjustment process.          

\appendix

\section{ISW propagation over a long canopy}\label{appA}

\begin{figure}
    \centerline{\includegraphics[width = 1.0\textwidth]{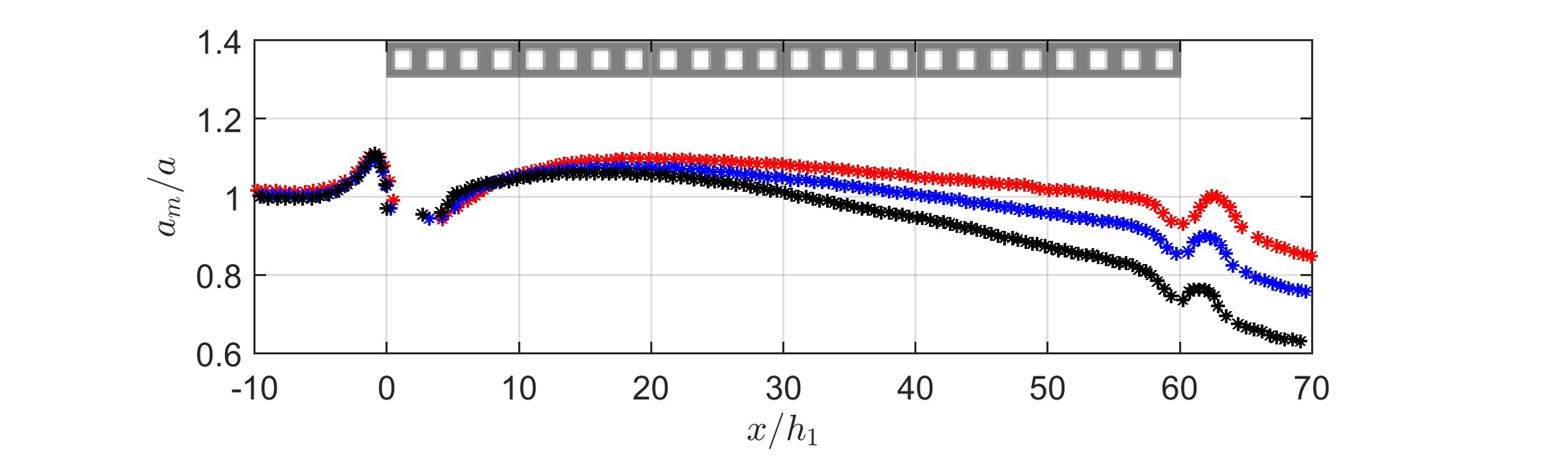}}
    \caption{Amplitude evolution in space for ISWs interacting with the dense canopy of $l_c/h_1 = 60$: A2(red), A1.5(blue), A1(black).}
    \label{fig:x_a_L12D}
\end{figure}

\begin{figure}
    \centerline{\includegraphics[width = 0.8\textwidth]{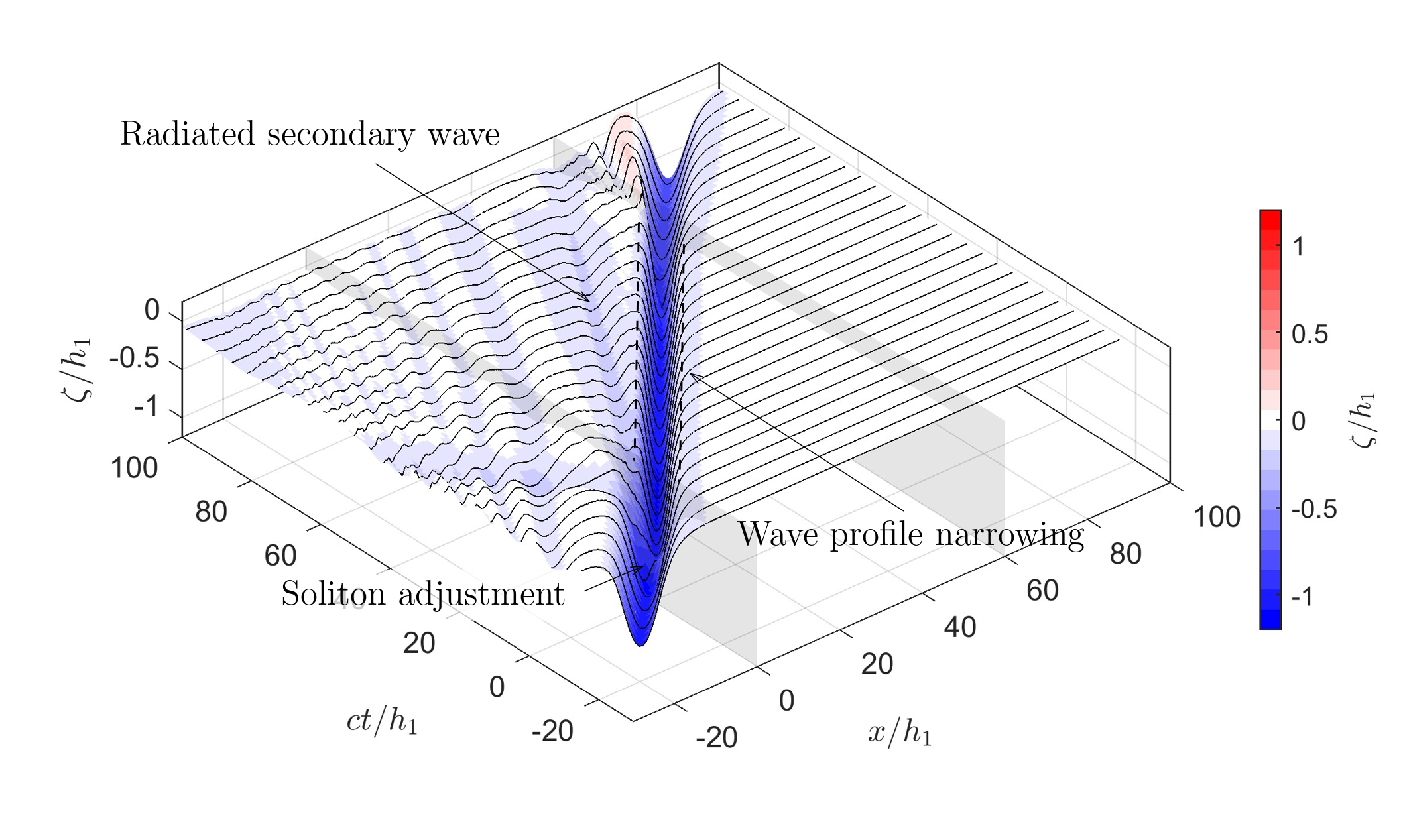}}
    \caption{Evolution of spatial wave profiles over time. The gray transparent planes at $x/h_1 = 0, 60$ indicate the leading edge and trailing edges of the dense canopy.}
    \label{fig:eta_3d_L12D}
\end{figure}

Despite the variation in canopy length of $l_c = 5h_1$ and  $l_c = 10h_1$ in the main discussion, the canopy length is less than half of the incident ISWs wavelength. Transient effects at the leading and trailing edges of the canopy can contaminate the development of a semi-steady, fully developed regime beneath the canopy. In order to isolate these transient effects, the interactions of the same ISWs with an extended dense canopy of $l_t = 60h_1$, which is roughly two times the incident ISWs wavelength, were performed inside a numerical flume of $3.2$ m. 

The amplitude evolution under the dense canopy of $l_c/h_1 = 60$ follows a pattern similar to that observed in Figure~\ref{fig:x_a}(e) under the dense canopy of $l_c/h_1 = 10$. The ISW undergoes an initial adjustment, then settles into a semi-steady propagation within $15<x/h_1<25$, as evidenced by the plateau trend in Figure~\ref{fig:x_a_L12D}. Beyond this interval, the ISW experiences sustained amplitude dissipation driven primarily by the drag at the bottom edge of the canopy until it exits the canopy structure. For the smaller-amplitude case ($a=1$ cm), the wave amplitude decreases by approximately 20\% at the end of the canopy, whereas the wave amplitude remains nearly unchanged for the larger-amplitude case ($a=2$ cm).
This behavior is consistent with the trend discussed in Section ~\ref{sec:isw_e_trans}, where smaller-amplitude ISWs were found to be more vulnerable to dissipation relative to larger-amplitude waves. The influence of the trailing-edge dynamics is effectively isolated from transient coupling effects when the canopy length is twice the ISW wavelength. A stronger jet induced by the flow separation downstream of the canopy, as shown in Figure~\ref{fig:vor_stp}(b1), unevenly accelerates the leading half of the soliton and produces an abrupt amplitude drop at $x/h_1=60$. The canopy length of twice the ISW wavelength also isolates the trailing edge effect from the transient coupling. The stronger jet induced by the flow separation downstream of the canopy, as shown in Figure~\ref{fig:vor_stp}(b1), sucts the leading half of the soliton out, which leads to the sudden amplitude drop right at $x/h_1=60$. In L1D and L2D, the amplitude drop caused by the flow separation cannot be clearly distinguished (Figure~\ref{fig:x_a}(c,e)). In L1D, it is coupled with the soliton adjustment phase, and it coincides with the onset of semi-steady propagation in L2D. Afterward, the vortex-driven steepening process reverses the descending trend locally, yielding a partial recovery of the wave amplitude downstream.

Figure~\ref{fig:eta_3d_L12D} presents a 3D view of the spatial wave profile evolution over time for an ISW of $a=2$ cm interacting with the dense canopy of $l_t = 60h_1$. Immediately after the soliton adjustment near the leading edge at $x/h_1=0$, shoaling and narrowing of the main soliton begin at the onset of the phase speed adjustment stage and continue until the canopy exit at $x/h_1=60$. A dispersive secondary wave is radiated from the narrowed transmitted soliton, identifiable as a light-blue stripe with a slower phase speed. This behavior resembles the ISW fission process described in the theoretical study of \citet{Grimshaw2008} for ISWs encountering a bottom step.      

\bibliographystyle{jfm}
\bibliography{jfm}

\end{document}